\documentclass[12pt]{article}

\catcode`\@=11
\@addtoreset{equation}{section}

\global\arraycolsep=1pt
\def\fnote#1#2{\begingroup\def\thefootnote{#1}\footnote{#2}\addtocounter
{footnote}{-1}\endgroup}

\usepackage{amsbsy,amssymb,latexsym,amsfonts}

\newcommand{\beq}{\begin{equation}}
\newcommand{\eeq}{\end{equation}}
\newcommand{\beqa}{\begin{eqnarray}}
\newcommand{\eeqa}{\end{eqnarray}}
\newcommand{\bR}{{\mathbb R}}

\newcommand{\CP}{{\mathcal P}}
\newcommand{\CN}{{\mathcal N}}
\newcommand{\CL}{{\mathcal L}}
\newcommand{\CD}{{\mathcal D}}
\newcommand{\CF}{{\mathcal F}}

\newcommand{\CV}{{\mathcal V}}
\newcommand{\CQ}{{\mathcal Q}}
\newcommand{\CS}{{\mathcal S}}

\newcommand\ad{{\rm ad}}

\newcommand\imag{{\rm Im}}
\newcommand\real{{\rm Re}}

\newcommand\intd{{\int\!d}{}}

\begin{document}
\begin{flushright}
OCU-PHYS-234\\
OIQP-05-11
\end{flushright}
\bigskip

\begin{center}
{\bf\Large
Partial Supersymmetry Breaking\\
and\\
$\CN=2$ $U(N_c)$ Gauge Model with Hypermultiplets\\
in Harmonic Superspace
}

\bigskip\bigskip
{
K. Fujiwara$^a$\fnote{$*$}{e-mail: \texttt{fujiwara@sci.osaka-cu.ac.jp}}
\quad, \quad
H. Itoyama$^a$\fnote{$\dag$}{e-mail: \texttt{itoyama@sci.osaka-cu.ac.jp}}
\quad and \quad
M. Sakaguchi$^b$\fnote{$\ddagger$}{e-mail: \texttt{makoto\_sakaguchi@pref.okayama.jp}}
}
\end{center}

\bigskip

\begin{center}
$^a$ {\it Department of Mathematics and Physics,
Graduate School of Science\\
Osaka City University\\
3-3-138, Sugimoto, Sumiyoshi-ku, Osaka, 558-8585, Japan }

\bigskip

$^b$ {\it Okayama Institute for Quantum Physics\\
1-9-1 Kyoyama, Okayama 700-0015, Japan}

\end{center}

\vfill

\begin{abstract}
We provide a manifestly $\CN=2$ supersymmetric formulation of
the
$\CN=2$ $U(N_c)$ gauge model
constructed in terms of $\CN=1$
superfields in hep-th/0409060.
The model is
composed of $\CN=2$ vector multiplets
in harmonic superspace
and can be viewed as
the $\CN=2$ $U(N_c)$ Yang-Mills effective action
equipped with the electric and magnetic Fayet-Iliopoulos terms.
We generalize this gauge model
to an $\CN=2$ $U(N_c)$ QCD model
by introducing 
$\CN=2$ hypermultiplets in harmonic superspace
which include both the fundamental representation of $U(N_c)$
and the adjoint representation of $U(N_c)$.
The effect of the magnetic Fayet-Iliopoulos term
is to shift the auxiliary field by an imaginary constant.
Examining vacua of the model,
we show that $\CN=2$ supersymmetry is spontaneously broken 
down to $\CN=1$.
\end{abstract}

\section{Introduction}
It is widely appreciated that
$\CN=2$ supersymmetry imposes a strong constraint
on the four-dimensional theory
but yet leaves rich physical ingredients.
For example, $\CN=2$ supersymmetric field theories
develop controllable quantum effects \cite{SW}.
Unconstrained $\CN=2$ superfields
provide a manifestly $\CN=2$ supersymmetric formulation of them.

There are two types of unconstrained $\CN=2$ vector superfields.
First type, developed in \cite{Mezincescu:1979af} for the abelian case
and in \cite{Howe:1983sr}\cite{Koller:1982hv} for the non-abelian case,
is  constructed on the usual $\CN=2$ 
superspace $\bR^{4|8}$ parametrized by
$(x^m,\theta^i,\bar\theta^i)$.
The second type is constructed on harmonic superspace
$\bR^{4|8}\times S^2$ 
developed in
\cite{HS1}
\cite{HS2}
\cite{HS3}
\cite{GIO1986}
(see \cite{HS} for introduction),
and 
parametrized by
\begin{eqnarray}
(x_A^m,\theta^\pm,\bar\theta^\pm,u_i^\pm)=
(x^m-2i\theta^{i}\sigma^m\bar\theta^{j}u_{(i}^+u_{j)}^-,\theta^iu_i^\pm,\bar\theta^iu_i^\pm,u_i^\pm)
\label{AB}
\end{eqnarray}
in the analytic basis.
The $S^2=SU(2)/U(1)$ is parametrized by harmonic variables $u^\pm_i$
\begin{eqnarray}
(u^+_i,u^-_i)\in SU(2)~,~~~
u^+_i=\varepsilon_{ij}u^{+j}~,~~~
u^{+i}u_i^-=1~,~~~
\overline{u^{+i}}=u^-_i~.
\end{eqnarray}
For  $\CN=2$ hypermultiplets,
harmonic superspace makes it possible to construct 
the
off-shell $\CN=2$ unconstrained  hypermultiplets,
the $q^+$- and $\omega$-hypermultiplets.
Thus harmonic superspace 
provides
a manifestly $\CN=2$ supersymmetric formulation of 
$\CN=2$ supersymmetric theories
in terms of off-shell $\CN=2$ unconstrained  superfields.
This property is very useful in the quantum calculations
\cite{LEEA in HS}
of supersymmetric models.
In the present paper, $\CN=2$ superfields in harmonic superspace
are utilized in the construction of $\CN=2$ $U(N_c)$ gauge model
coupled with $\CN=2$ hypermultiplets.

In \cite{FIS1},
the $\CN=2$ $U(N_c)$ gauge model is constructed in terms of 
$\CN=1$ vector and chiral superfields
in the adjoint representation of the gauge group $U(N_c)$.
This model
is a non-abelian generalization of
the $U(1)$ gauge model
with abelian constrained $\CN=2$ vector superfields \cite{N=2 constraint}
constructed by
Antoniadis, Partouche and Taylor (APT) in \cite{APT},
and can be regarded as
$\CN=2$ $U(N_c)$ Yang-Mills (YM) low-energy
effective action (LEEA)
specified by a holomorphic function $\CF$ 
and equipped 
with the electric and magnetic Fayet-Iliopoulos (FI) terms.
In \cite{FIS1}
\cite{Fujiwara:2004sz}
\cite{Fujiwara:2005hj},
the vacua of the model are examined.
The $\CN=2$ and $U(N_c)$ symmetry are
spontaneously broken to
$\CN=1$ and $\displaystyle\prod_nU(N_n)$ symmetry with 
$\displaystyle\sum_nN_n=N_c$,
respectively.
The associated Nambu-Goldstone (NG) fermion
is shown to be provided by the overall $U(1)$ part in $U(N_c)$.
In addition, the spectrum on the vacua is
completely clarified in \cite{Fujiwara:2005hj}.\fnote{$\sharp$}{
This series of works
\cite{FIS1}\cite{Fujiwara:2004sz}\cite{Fujiwara:2005hj}
is based on $\CN=1$ superspace
and construction of most general $\CN=2$
Lagrangian based on special K\"ahler geometry
\cite{N=2 sugra}
which were developed after tensor calculus
\cite{N=2 sugra 2}.
}

The key for the partial supersymmetry breaking is
the modification of the local version of the extended supersymmetry
algebra by an additional spacetime independent term
\cite{Lopuszanski:1978df,central charge}. 
This modification is caused by 
the magnetic FI term.
An important observation made in the study of the APT model
\cite{APT}
\cite{central charge}
\cite{Partouche:1996yp}
\cite{Ivanov:1997mt}
is that
the effect of the magnetic FI term is 
to shift the auxiliary field in the $\CN=2$ vector superfield
by an imaginary constant
(while that of the electric FI term 
is to shift the symplectic dual auxiliary field
by an imaginary constant).
This observation is shown to be useful
in this paper
in the construction of the magnetic FI term
for $\CN=2$ $U(N_c)$ gauge model with/without hypermultiplets.

In this paper, 
we provide a manifestly $\CN=2$ supersymmetric formulation of
the
$\CN=2$ $U(N_c)$ gauge model given in
\cite{FIS1}.
For this purpose we employ $\CN=2$ vector multiplets $V^{++}$
in harmonic superspace.
The magnetic FI term
is introduced so as to shift the auxiliary field in $V^{++}$ 
by an imaginary constant.
This causes  $\CN=2$ supersymmetry to be broken spontaneously
to $\CN=1$.
In addition,
we generalize this gauge model
to an $\CN=2$ $U(N_c)$ QCD model coupled with 
$\CN=2$  hypermultiplets,
$q^+$ and $\omega$,
in harmonic superspace
which include both
the fundamental representation of $U(N_c)$
and 
the adjoint representation of $U(N_c)$.
We determine the form of the magnetic FI term
such that it shifts the auxiliary field in $V^{++}$  by an imaginary constant.
Examining vacua of the model,
we show that this model also describes partial supersymmetry breaking.
It should be noted that 
the magnetic FI term of the $U(N_c)$ gauge model coupled
with hypermultiplets in the adjoint representation 
is the same as that of the $U(N_c)$
gauge model without hypermultiplets.
However in presence of hypermultiplets in the fundamental representation
the magnetic FI term develops an additional term.
This additional term overcomes the
difficulty
\cite{Partouche:1996yp} 
\cite{Ivanov:1997mt} in coupling fundamental hypermultiplets to the APT model.

This paper is organized as follows.
In section 2, 
we introduce $\CN=2$ vector multiplets $V^{++}$
in harmonic superspace and
construct an $\CN=2$ $U(N_c)$ gauge model
equipped with the electric and magnetic FI terms.
The vacua of the model are examined in section 3.
We show that the model describes  spontaneous 
partial supersymmetry breaking.
Introducing $\CN=2$ hypermultiplets, $q^+$ and $\omega$,
 in harmonic superspace
we generalize the model to the $\CN=2$ $U(N_c)$ QCD model equipped with
the electric and magnetic FI terms in section 4.
We introduce the magnetic FI term so as to shift the auxiliary field
by an imaginary constant.
In section 5, we show that
owing to this property,
$\CN=2$ supersymmetry is broken down to
$\CN=1$ spontaneously. 
The supersymmetry transformation law of the components
in the vector multiplet $V^{++}$ and
the hypermultiplets, $q^+$ and $\omega$, is found in appendix B.
In this paper,
we follow the notation for harmonic superspace
given in \cite{HS} (see appendix A)
and one for spinors given in \cite{WB}.

\section{$\CN=2$ $U(N_c)$ gauge model}

We introduce a set of  $\CN=2$ vector superfields $V^{++}=V^{++a}t_a$ where
$N_c\times N_c$ hermitian matrices $t_a$ ($a=0,1,...,N_c^2-1$)
generate 
$u(N_c)$, $[t_a,t_b]=if^a_{bc}t_a$, 
and $t_0$ represents the overall $u(1)$ generator.
$V^{++}$ is the analytic superfield satisfying
$D^+V^{++}=\bar D^+V^{++}=0$.
In the analytic basis $(x_A,\theta^\pm,\bar\theta^\pm,u^\pm_i)$ in
(\ref{AB}),
$D^\pm$ and $\bar D^\pm$ are given as
\begin{eqnarray}
&&
D^+_\alpha=\frac{\partial}{\partial\theta^{-\alpha}}~,~~~
D^-_\alpha=-\frac{\partial}{\partial\theta^{+\alpha}}
+2i(\sigma^m\bar\theta^{-})_\alpha\frac{\partial}{\partial x_A^m}~,
\nonumber\\&&
\bar D^+_{\dot \alpha}=\frac{\partial}{\partial\bar\theta^{-\dot\alpha}}~,~~~
\bar D^-_{\dot \alpha}=
-\frac{\partial}{\partial\bar\theta^{+\dot\alpha}}
-2i(\theta^{-}\sigma^m)_{\dot \alpha}\frac{\partial}{\partial x_A^m}~,~~~
\end{eqnarray}
and thus $V^{++}$
is a superfield
in analytic subspace
$(x_A,\theta^+,\bar\theta^+,u^\pm_i)$~.
In the Wess-Zumino (WZ) gauge
$V^{++}$ is given as
\begin{eqnarray}
V^{++}&=&-2i\theta^+\sigma^m\bar\theta^+v_m(x_A)
-i\sqrt{2}(\theta^+)^2\bar\phi(x_A)
+i\sqrt{2}(\bar\theta^+)^2\phi(x_A)
\nonumber\\&&
+4(\bar\theta^+)^2\theta^+\lambda^i(x_A)u_i^-
-4(\theta^+)^2\bar\theta^+\bar\lambda^i(x_A)u_i^-
+3(\theta^+)^2(\bar\theta^+)^2D^{ij}(x_A)u_i^-u_j^-~,
\label{V++}
\end{eqnarray}
where $v_m$, $\phi$, $\lambda^i$ and $D^{ij}$
are vector, complex scalar, $SU(2)$ doublet Weyl spinor and 
auxiliary field, respectively.
$D^{ij}$ is symmetric with respect to $(i,j)$ so that
$D^i{}_{j}=\varepsilon_{jk}D^{ik}$
is an $SU(2)$ matrix, $D^i{}_{j}=iD^A(\tau_A)^i{}_{j}$.
The reality $V^{++}=\widetilde{V^{++}}$,
where 
the tilde
``$\widetilde{~~~}$'' 
means the analyticity preserving 
conjugation \cite{HS} (see appendix A), implies that
$D^{ij}=\bar D^{ij}$
because
$\overline{D^{ij}u_i^-u_j^-}=\bar D_{ij}u^{-i}u^{-j}=
\bar D^{ij}u^{-}_iu^{-}_j$,
and that the $D^A$ is a real three vector
$
\overline{D^A}=D^A~.
$

The field strength $\overline{W}$ is given by
\begin{eqnarray}
\overline{W}&=&
-\frac{1}{4}(D^+)^2
\sum_{n=1}^{\infty}
\intd v_1\cdots dv_n(-i)^{n+1}
\frac{V^{++}(v_1)\cdots V^{++}(v_n)}
{(u^+v_1^+)(v_1^+v_2^+)\cdots(v_n^+u^+)}~.
\end{eqnarray}
It is straightforward to derive
the following expression
by using formulas given in \cite{HS}(see also \cite{HS2}):
\begin{eqnarray}
\overline{W}&=&
-i\sqrt{2}\bar\phi
-2\bar\theta^i\bar\lambda_i
+\bar\theta^i\bar\theta^jD_{ij}
+\bar\theta^i\bar\sigma^{mn}\bar\theta_i\, v_{mn}
+\frac{4}{3}i(\bar\theta^i\bar\theta^j)\,\CD_m\lambda_i\sigma^m\bar\theta_j
-\frac{2}{3}\sqrt{2}(\bar\theta^i\bar\theta^j)[\phi, \bar\theta_i\bar\lambda_j]
\nonumber\\&&
+i\sqrt{2}(\bar\theta)^4\eta^{mn}\CD_{m}\CD_n\phi
+i(\bar\theta)^4\,\varepsilon_{ij}[\lambda^i, \lambda^j]
+i\sqrt{2}(\bar\theta)^4[\phi,[\phi,\bar\phi]]
{-2i\bar\theta^+\bar\theta^-[\phi,\bar\phi]
+\cdots}
\label{W}
\end{eqnarray}
where
\begin{eqnarray}
v_{mn}&=&\partial_mv_n-\partial_nv_m+i[v_m,v_n]~,~~~
\nonumber\\
\CD_m\phi&=&\partial_m\phi+i[v_m,\phi]~,~~~
\CD_m\lambda=\partial_m\lambda+i[v_m,\lambda]~
\end{eqnarray}
and ellipsis represents terms which do not contribute to the action.

We consider the action
\begin{eqnarray}
S_V=-\frac{i}{4}\intd^4x\left[
( D)^4{\CF}({W})
-
(\bar D)^4\overline{\CF}(\overline{W})
\right]
~,
\label{SV:superfield}
\end{eqnarray}
where $\CF$ is an analytic trace function of 
${W}={W}^at_a$ and
$(D)^4=\frac{1}{16}(D^+)^2(D^-)^2$.
The $S_V$ can be regarded as the leading contribution to the
$\CN=2$ YM LEEA.
The action (\ref{SV:superfield}) reduces in component fields to
\begin{eqnarray}
S_V=
\intd^4x\Big[&&
-g
_{ab}
\CD^m\phi^a \CD_m\bar\phi^b
-\frac{1}{2}\overline{\CF}_{ab}|\bar\lambda^{ai}\bar\sigma^m\CD_m\lambda^b_i
+\frac{1}{2}{\CF}_{ab}|\lambda^{ai}\sigma^m\CD_m\bar\lambda^{b}_i
\nonumber\\&&
-\frac{1}{4}
g
_{ab}v_{mn}^av^{b\,mn}
-\frac{1}{8}\real\CF_{ab}|\varepsilon^{mnpq}v^a_{mn}v^b_{pq}
+{\frac{1}{2}}
g
_{ab}\,f^a_{cd}\bar\phi^c\phi^d\,
f^b_{ef}\bar\phi^e\phi^f
\nonumber\\&&
+\frac{1}{4}
g
_{ab}D^{a\,ij}D_{ij}^b
+\frac{i}{4}
\CF_{abc}|
\lambda^{ai}\lambda^{bj}
D_{ij}^c
-\frac{i}{4}\overline{\CF}_{abc}|
\bar\lambda^{ai}\bar\lambda^{bj}
D_{ij}^c
\nonumber\\&&
-\frac{i}{4}(
\CF_{abc}|\lambda^{ai}\sigma^{mn}\lambda^{b}_i 
-
\overline{\CF}_{abc}|\bar\lambda^{ai}\bar\sigma^{mn}\bar\lambda^b_i 
)v^c_{mn}
\nonumber\\&&
+\frac{1}{2}
g_{ab}
\left[
\bar\lambda^{ai}f^b_{cd}(i\sqrt{2}\phi^c)\bar\lambda^d_i
+
\lambda^{ai}f^b_{cd}(-i\sqrt{2}\bar\phi^c)\lambda^d_i
\right]
\nonumber\\&&
-\frac{i}{12}\CF_{abcd}|(\lambda^{ai}\lambda^{bj})(\lambda^c_i\lambda^d_j)
+\frac{i}{12}\overline{\CF}_{abcd}|
(\bar\lambda^{ai}\bar\lambda^{bj})(\bar\lambda^c_i\bar\lambda^d_j)
~~\Big]~,~~
\label{SV:component}
\end{eqnarray}
where $\CF_{a_1\cdots a_n}\equiv 
{\partial^n\CF}/{\partial W^{a_1}\cdots\partial W^{a_n} }$ 
and $g_{ab}\equiv \imag\CF_{ab}|$.
$\CF_{ab\cdots}|$ represents $\CF_{ab\cdots}$ evaluated at 
$\theta^\pm=\bar\theta^\pm=0$.
We have used the relation
\begin{eqnarray}
f^c_{ba}\CF_c|=-f^c_{bd}(i\sqrt{2}\phi^d)\CF_{ca}|
\end{eqnarray}
which follows from the fact that
$\phi^a$ and 
$\CF_a|$ transform as adjoint under $U(N_c)$.

\subsection{electric \& magnetic FI terms}

We
introduce the electric FI term \cite{FI}
\begin{eqnarray}
S_{e}&=&
\intd ud\zeta^{(-4)}{\rm tr}\,(\Xi^{++}
V^{++})
+c.c.
=
\intd^4x \xi^{ij}
D_{ij}^0
+c.c.
~,
\label{Se}
\end{eqnarray}
where $d\zeta^{(-4)}=d^4x_Ad^4\theta^+$
and $\Xi^{++}=\xi^{ij}u_i^+u_j^+$ is the electric FI parameter.
This term develops a constant imaginary part in the dual auxiliary field $D_{D}^{aij}$
of $W_{D}^a\equiv\CF_a$.
To see this we 
derive the equation of motion for $V^{++}$ from $S_V+S_e$.
Rewriting $S_V$ as an integral over the analytic subspace
as was done in \cite{Zupnik} for the bare theory
in which $\CF$ is quadratic,
and then varying with respect to $V^{++}$,
we obtain
\begin{eqnarray}
\frac{i}{16}(D^+)^2\CF_a+\Xi^{++}\delta_a^0+c.c=0~.
\end{eqnarray}
This may be rewritten as
\begin{eqnarray}
(D^+)^2(\CF_a+8i\xi^{ij}\theta_i\theta_j\delta^a_0)
-
(\bar D^+)^2(\overline{\CF}_a-8i\bar\xi^{ij}\bar\theta_i\bar\theta_j\delta^a_0)
=0~.
\end{eqnarray}
Because $\CF_a=W_D^a=\theta^i\theta^jD_{Dij}^a+\cdots$,
the effect of the electric FI term is to shift
the dual auxiliary field
by an imaginary constant,
  $D_D^{aij}\to D_D^{aij}+8i\xi^{ij}\delta^a_0$. 

Next we introduce the magnetic FI term of the form
\begin{eqnarray}
S_m^{\rm YM}&=&
\intd^4x\,
(D)^4
\xi^{ij}_D\theta_i\theta_j\left(
{\CF}_0
+\frac{1}{2}\CF_{00}4i\xi^{kl}_D\theta_k\theta_l
\right)
+c.c.\nonumber\\
&=&
\intd^4x\, \left[
\xi^{ij}_D(
{\CF}_{0a}|D_{ij}^a
-{\CF}_{0ab}|\lambda^a_i\lambda_j^b
)
+
2i\CF_{00}|\xi^{ij}_D\xi_{Dij}
)
\right]
+c.c.~.
\label{mag.FI:YM}
\end{eqnarray}
We note that
this term reduces to $S_e+const.$ for the bare theory.
The effect of this term is to shift the auxiliary field $D^{aij}$
by an imaginary constant.
This can be seen as follows.
Observe that $S_m^{\rm YM}$ and  the $D$-dependent terms in $S_V$
may be rewritten as
\begin{eqnarray}
-\frac{i}{4}\intd^4x(D^4)&\Big[&
\CF_a|\theta^i\theta^j(\boldsymbol{D}^a_{ij})
\nonumber\\&&
+\frac{1}{2}\CF_{ab}|\left(
\theta^i\theta^j\boldsymbol{D}^a_{ij}\,\theta^k\theta^l\boldsymbol{D}^b_{kl}
+2\theta^i\theta^j\boldsymbol{D}^a_{ij}(-2\theta^k\lambda_k^b+\theta^k\sigma^{mn}\theta_kv_{mn}^b)
\right)
\nonumber\\&&
+\frac{1}{6}\CF_{abc}|3\theta^i\theta^j\boldsymbol{D}_{ij}^a
 (-2\theta^l\lambda_k^b)(-2\theta^l\lambda_l^c)
\Big]+c.c.
\end{eqnarray}
where
\begin{eqnarray}
\boldsymbol{D}^{aij}=D^{aij}+4i\xi^{ij}_D\delta^a_0~,~~~
\bar{\boldsymbol{D}}^{aij}=D^{aij}-4i\bar\xi^{ij}_D\delta^a_0~.
\label{bold D}
\end{eqnarray}
This implies that
\begin{eqnarray}
S_V+S_m^{\rm YM}=-\frac{i}{4}\intd^4x(D^4) \CF(\hat W)
+c.c.
\end{eqnarray}
where $\hat W$ is $W$ with the replacement $D^{aij}\to \boldsymbol{D}^{aij}$
(similarly $\overline{\hat W}$ is $\overline{W}$ with the replacement 
$D^{aij}\to \bar{\boldsymbol{D}} ^{aij}$).
We note that
due to this effect the supersymmetry transformation law 
$\delta_\eta\lambda^{ai}=(D^a)^i{}_j\eta^j+\cdots$
(see appendix B)
changes to
\begin{eqnarray}
\delta_\eta\lambda^{ai}=(\boldsymbol{D}^a)^i{}_j\eta^j+\cdots~,~~~
\delta_\eta\bar\lambda^{ai}=-(\bar{\boldsymbol{D}}^a)^i{}_j\bar\eta^j+\cdots~.
\label{susy tfn}
\end{eqnarray}

Gathering all together, the total action 
for the $U(N_c)$ gauge model
is
\begin{eqnarray}
S_{\rm YM}=S_V+S_e+S_m^{\rm YM}~.
\label{S_YM}
\end{eqnarray}
The terms including the auxiliary field $D^a_{ij}$
\begin{eqnarray}
\intd^4x&\Big[&
\frac{1}{4}g_{ab}D^{aij}D^b_{ij}
+(\xi^{ij}+\bar\xi^{ij})D_{ij}^0
+(\xi^{ij}_D\CF_{0a}|+\bar\xi^{ij}_D\overline{\CF}_{0a}|)D_{ij}^a
\nonumber\\&&
+\frac{i}{4}\CF_{abc}|\lambda^{ai}\lambda^{bj}D_{ij}^c
-\frac{i}{4}\overline{\CF}_{abc}|\bar\lambda^{ai}\bar\lambda^{bj}D_{ij}^c
\Big]~
\end{eqnarray}
lead to
\begin{eqnarray}
D^{aij}=-2g^{ab}\Big[
(\xi^{ij}+\bar\xi^{ij})\delta_b^0
+\xi^{ij}_D\CF_{0b}|+\bar\xi^{ij}_D\overline{\CF}_{0b}|
+\frac{i}{4}\CF_{bcd}|\lambda^{ci}\lambda^{dj}
-\frac{i}{4}\overline{\CF}_{bcd}|\bar\lambda^{ci}\bar\lambda^{dj}
\Big]~.~~~
\end{eqnarray}
By eliminating the auxiliary field
using this equation,
the action $S_{\rm YM}$ becomes
\begin{eqnarray}
S_{\rm YM}'&=&\intd^4x\Big[
\CL_{\rm kin}+\CL_{\rm pot}+\CL_{\rm Pauli}+\CL_{\rm mass}+\CL_{\rm 4\,fermi}
\Big]
\label{action:YM'}
\end{eqnarray}
where
\begin{eqnarray}
\CL_{\rm kin}&=&
-g_{ab}\CD^m\phi^a \CD_m\bar\phi^b
-\frac{1}{2}\overline{\CF}_{ab}|\bar\lambda^{ai}\bar\sigma^m\CD_m\lambda^b_i
-\frac{1}{2}{\CF}_{ab}|\lambda^{ai}\sigma^m\CD_m\bar\lambda^{b}_i
\nonumber\\&&
-\frac{1}{4}g
_{ab}v_{mn}^av^{b\,mn}
-\frac{1}{8}\real\CF_{ab}|\varepsilon^{mnpq}v^a_{mn}v^b_{pq}~,
\label{YM:kin}\\
\CL_{\rm pot}&=&
-g_{ab}\CP^a\CP^b
-\frac{1}{4}g^{ab}
{D}^{aij}|~\bar{{D}}^a_{ij}|
+2i\xi_D^{ij}\xi_{Dij}\CF_{00}|
-2i\bar\xi_D^{ij}\bar\xi_{Dij}\overline{\CF}_{00}|
\nonumber\\&=&
-g_{ab}\CP^a\CP^b
-\frac{1}{4}g^{ab}
\boldsymbol{D}^{aij}|~\bar{\boldsymbol{D}}^a_{ij}|
-2i(\xi^{ij}+\bar\xi^{ij})(\xi_{Dij}-\bar\xi_{Dij})~,
\label{YM:pot}\\
\CL_{\rm Pauli}&=&
-\frac{i}{4}(
\CF_{abc}|\lambda^{ai}\sigma^{mn}\lambda^{b}_i 
-
\overline{\CF}_{abc}|\bar\lambda^{ai}\bar\sigma^{mn}\bar\lambda^b_i 
)v^c_{mn}~,
\label{YM:pauli}\\
\CL_{\rm mass}&=&
+\frac{1}{2}g_{ab}
\left[
\bar\lambda^{ai}f^b_{cd}(i\sqrt{2}\phi^c)\bar\lambda^d_i
+
\lambda^{ai}f^b_{cd}(-i\sqrt{2}\bar\phi^c)\lambda^d_i
\right]
\nonumber\\&&
-\frac{i}{2}g^{ab}\left((\xi^{ij}+\bar\xi^{ij})\delta_a^0
+\xi_D^{ij}\CF_{0a}|
+\bar\xi_D^{ij}\overline{\CF}_{0a}|\right)
(\CF_{bcd}|\lambda^c_i\lambda^d_j-\overline{\CF}_{bcd}|\bar\lambda^c_i\bar\lambda^d_j)
\nonumber\\&&
-\bar\xi^{ij}_D
\overline{\CF}_{0ab}|\bar\lambda^a_i\bar\lambda_j^b
-\xi^{ij}_D{\CF}_{0ab}|\lambda^a_i\lambda_j^b
~,
\nonumber\\&=&
+\frac{1}{2}g_{ab}
\left[
\bar\lambda^{ai}f^b_{cd}(i\sqrt{2}\phi^c)\bar\lambda^d_i
+
\lambda^{ai}f^b_{cd}(-i\sqrt{2}\bar\phi^c)\lambda^d_i
\right]
\nonumber\\&&
+\frac{i}{4}\boldsymbol{D}^{aij}|~\CF_{abc}|\lambda^b_i\lambda^c_j
-\frac{i}{4}\bar{\boldsymbol{D}}^{aij}|~\overline{\CF}_{abc}|\bar\lambda^b_i\bar\lambda^c_j
~,
\label{YM:mass}\\
\CL_{\rm 4\,fermi}&=&
-\frac{i}{12}\CF_{abcd}|(\lambda^{ai}\lambda^{bj})(\lambda^c_i\lambda^d_j)
+\frac{i}{12}\overline{\CF}_{abcd}|
(\bar\lambda^{ai}\bar\lambda^{bj})(\bar\lambda^c_i\bar\lambda^d_j)
\nonumber\\&&
+\frac{1}{16}g^{ab}
(\CF_{acd}|\lambda^{ci}\lambda^{dj}
-\overline{\CF}_{acd}|\bar\lambda^{ci}\bar\lambda^{dj})
(\CF_{bef}|\lambda^{e}_i\lambda^{f}_j
-\overline{\CF}_{acd}|\bar\lambda^{e}_i\bar\lambda^{f}_j)~,
\label{YM:4 fermi}
\end{eqnarray}
and
\begin{eqnarray}
{\sqrt{2}}\CP^a&=&
-if^a_{bc}\bar\phi^{b}\phi^c
=-ik_b{}^a\bar\phi^b
=+ik^*_b{}^a\phi^b
~,\label{P}\\
D^{aij}|&=&-2g^{ab}\Big[
(\xi^{ij}+\bar\xi^{ij})\delta_b^0
+\xi^{ij}_D\CF_{0b}|+\bar\xi^{ij}_D\overline{\CF}_{0b}|
\Big]~,\\
\boldsymbol{D}^{aij}|&=&D^{aij}|+4i\xi^{ij}_D\delta^a_0
=
-2g^{ab}\Big[
(\xi^{ij}+\bar\xi^{ij})\delta_b^0
+(\xi^{ij}_D+\bar\xi^{ij}_D)\overline{\CF}_{0b}|
\Big]~.
\label{bold D:YM}
\end{eqnarray}
$\CP_a$ is the Killing potential for the Killing vector $k_a=k_a{}^b\partial_b$
which generates $U(N_c)$ isometry of 
 the special K\"ahler geometry.

\section{Vacua of $\CN=2$ $U(N_c)$ gauge model}
We examine vacua of the model defined by $S_{\rm YM}=S_V+S_e+S_m^{\rm YM}$.
The scalar potential $\CV=-\CL_{\rm pot}$ is 
\begin{eqnarray}
\CV&=&
\frac{1}{4}g_{ab}
\boldsymbol{D}^{aij}|~\bar{\boldsymbol{D}}^b_{ij}|
+g_{ab}\CP^a\CP^b
+2i(\xi^{ij}+\bar\xi^{ij})(\xi_{Dij}-\bar\xi_{Dij})
~.
\end{eqnarray}
In the three-vector notation,
\begin{eqnarray}
&&\xi^i{}_{j}=i\xi^A(\tau_A)^i{}_{j}~,~~~
\xi_{D}{}^i{}_{j}=i\xi_{D}^A(\tau_A)^i{}_{j}~,~~~
\boldsymbol{D}^{ai}{}_{j}=i\boldsymbol{D}^{aA}(\tau_A)^i{}_{j}~,~~~
A=1,2,3,\nonumber\\
&&\boldsymbol{D}^{aij}\bar{\boldsymbol{D}}^b_{ij}=
-\boldsymbol{D}^{ai}{}_j\bar{\boldsymbol{D}}^{bj}{}_i=
\boldsymbol{D}^{aA}\bar{\boldsymbol{D}}^{bB}
{\rm tr}(\tau_A\tau_B)=
2\sum_A\boldsymbol{D}^{aA}\bar{\boldsymbol{D}}^{bA}
\end{eqnarray}
this is written as
\begin{eqnarray}
\CV&=&
\frac{1}{2}g_{ab}
\boldsymbol{D}^{aA}|~\bar{\boldsymbol{D}}^{bA}|
+g_{ab}\CP^a\CP^b
+4i(\xi^{A}+\bar\xi^{A})(\xi_{D}^A-\bar\xi_{D}^A)
\nonumber\\
&=&2g^{ab}\left|\left[
(\xi^{A}+\bar\xi^{A})\delta_a^0
+(\xi^{A}_D+\bar\xi^{A}_D)\overline{\CF}_{0a}|
\right]
\right|^2
+g_{ab}\CP^a\CP^b
+4i(\xi^{A}+\bar\xi^{A})(\xi_{D}^A-\bar\xi_{D}^A)~.
\label{V: YM}
\end{eqnarray}
The last term is a constant.
We demand positive definiteness of $g_{ab}$.
In order to determine the vacuum,
we examine ${\partial\CV}/{\partial(W^a|)}=0$.
The second term $g_{ab}\CP^a\CP^b$ in $\CV$ tells us that
$\langle\phi^r\rangle=0$ where $t_r$ represent non-Cartan generators
and $\langle *\rangle$ means the vacuum expectation value (vev) of $*$.
The vacuum condition is
\cite{Fujiwara:2004sz}
\begin{eqnarray}
\langle
\partial_a\CV
\rangle&=
\frac{i}{4}
\langle
\CF_{abc}|~
\boldsymbol{D}^{bA}\boldsymbol{D}^{cA}
\rangle=0~
\end{eqnarray}
where we note $\langle\boldsymbol{D}^{cA}|\rangle=\langle\boldsymbol{D}^{cA}\rangle$
because fermions do not acquire vevs.
Let us examine the case with
\begin{eqnarray}
\CF=\sum_n\frac{c_n}{n!}{\rm tr}(W)^n
\end{eqnarray}
for concreteness.
Let $E_{\underline{\rm i}\underline{\rm j}}$, 
$\underline{{\rm i}}=\underline{1},...,\underline{N_c}$ 
be the fundamental matrix which has $1$ at 
the $(\rm \underline{\rm i}, \underline{\rm j})$-component
and $0$ otherwise.
Cartan generators may be written as 
$t_{\underline{\rm i}}=
E_{\underline{\rm i}\underline{\rm i}}$.
We have $\langle\partial_r\CV\rangle=0$, because
$\CF_{r\underline{{\rm i}}\underline{{\rm i}}}=\langle\boldsymbol{D}^r\rangle=0$.
Noting that 
the points specified by 
$\langle\CF_{\underline{{\rm i}}\underline{{\rm i}}\underline{{\rm i}}}\rangle=0$
correspond to unstable vacua,
we derive the vacuum condition
\begin{eqnarray}
\sum_A
\langle\boldsymbol{D}^{\underline{\rm i}A}
\boldsymbol{D}^{\underline{\rm i}A}\rangle=0~,~~~
\underline{\rm i}=\underline{1},...,\underline{N_c}~.
\label{vacuum condition}
\end{eqnarray}
This is solved by
\begin{eqnarray}
\langle\phi\rangle=
{\rm diag}(\stackrel{N_1}{\overbrace{a_1,\cdots,a_1}},
\stackrel{N_2}{\overbrace{a_2,\cdots,a_2}},\cdots)~,
\end{eqnarray}
which means that the gauge symmetry $U(N_c)$
is broken down  to $\displaystyle\prod_n U(N_n)$ with $\sum N_n=N_c$.

On the other hand,
the supersymmetry transformation of $\lambda^{ai}$
(\ref{susy tfn}) reduces on the vacuum to
\begin{eqnarray}
\langle\delta_{\eta}\lambda^{\underline{\rm i}i}\rangle
=\langle \boldsymbol{D}^{\underline{\rm i}i}{}_j\rangle\eta^j
=i\langle \boldsymbol{D}^{\underline{\rm i}A}\rangle (\tau_A)^i{}_j\eta^j
~~,~~~~
\langle\delta_{\eta}\lambda^{ri}\rangle
=\langle \boldsymbol{D}^{ri}{}_j\rangle\eta^j
=0~.
\end{eqnarray}
A necessary condition for the partial supersymmetry breaking
is
\begin{eqnarray}
\langle\det \boldsymbol{D}^{\underline{\rm i}i}{}_j\rangle=
2\sum_A\langle
\boldsymbol{D}^{\underline{\rm i}A}\boldsymbol{D}^{\underline{\rm i}A}
\rangle
=0~
\label{determinant}
\end{eqnarray}
for some $\underline{\rm i}$.
This is obviously satisfied by the vacua (\ref{vacuum condition}).
Let us look at the mass term for $\lambda^{\underline{\rm i}}_i$
\begin{eqnarray}
-\frac{i}{4}\langle \CF_{\underline{\rm i}\underline{\rm i}\underline{\rm i}}|
\rangle
\lambda^{\underline{\rm i}}_i
\langle\boldsymbol{D}^{\underline{\rm i}}\rangle^i{}_j
\lambda^{\underline{\rm i}j}
=-\frac{i}{4}\langle 
\CF_{\underline{\rm i}\underline{\rm i}\underline{\rm i}}|
\rangle\langle\boldsymbol{D}^{\underline{\rm i}A}\rangle
\lambda^{\underline{\rm i}i}
(\tau_2\tau_A)_{ij}
\lambda^{\underline{\rm i}j}
~.
\label{lambda: mass}
\end{eqnarray}
Because of
 (\ref{determinant}) ,
a half of the fermions $\lambda^{\underline{\rm i}}_i$, $i=1,2$, is massless.
In fact, 
diagonalizing the mass matrix by a matrix $U^i{}_j$,
we see that
the supersymmetry transformation
of the massless combination of $\lambda^{\underline{\rm i}i}$,
say $U^1{}_j\lambda^{\underline{\rm i}j}$,
is non-trivial.
In the ordinary hermitian matrices $t_a$,
$a=0,\cdots, N_c^2-1$,
of $u(N_c)$, this means that
$U^1{}_j\lambda^{0j}$ is massless and $\delta_\eta (U^1{}_j\lambda^{0j})\neq 0$.
Thus the Nambu-Goldstone fermion for partial supersymmetry breaking
lies in the overall $U(1)$ part of the massless combination of 
$\lambda^{\underline{\rm i}i}$, $U^1{}_j\lambda^{0j}$.

Let us make a comment on the relation between the present construction
and that of 
\cite{FIS1}.
The vacuum condition means that
\begin{eqnarray}
&&\left(
2(\xi^A+\bar\xi^A)+(\xi_D^A+\bar\xi_D^A)
\langle\real\CF_{\underline{\rm i}\underline{\rm i}}|\rangle
\right)^2=
(\xi_D^A+\bar\xi_D^A)^2\langle
g_{\underline{\rm i}\underline{\rm i}}\rangle^2~,
\label{vacuum condition 1}
\\&&
\left(
2(\xi^A+\bar\xi^A)+(\xi_D^A+\bar\xi_D^A)\langle
\real\CF_{\underline{\rm i}\underline{\rm i}}|\rangle
\right)
(\xi_D^A+\bar\xi_D^A)\langle g
_{\underline{\rm i}\underline{\rm i}}\rangle
=0~.\label{vacuum condition 2}
\end{eqnarray}
By using $SU(2)$, 
we may choose $(\xi_D^A+\bar\xi_D^A) = (0,-m,0)$ with a real constant $m$
without loss of generality.
Then (\ref{vacuum condition 2}) implies
\begin{eqnarray}
\langle\real\CF_{\underline{\rm i}\underline{\rm i}}\rangle
=\frac{2}{m}(\xi^2+\bar\xi^2)
\equiv -2\frac{e}{m}~
\end{eqnarray}
with a real constant $e$.
Substituting this into (\ref{vacuum condition 1}) we find
\begin{eqnarray}
\langle g_{\underline{\rm i}\underline{\rm i}}\rangle
=\pm\frac{2}{m}
\sqrt{(\xi^1+\bar\xi^1)^2+(\xi^3+\bar\xi^3)^2}~.
\end{eqnarray}
By using $U(1)$, residual rotational symmetry along $2$-axis,
we may choose $(\xi^1+\bar\xi^1)=0$ and $(\xi^3+\bar\xi^3)=\xi$
with a real constant $\xi$
without loss of generality,
so that
\begin{eqnarray}
\langle g_{\underline{\rm i}\underline{\rm i}}\rangle
=\mp 2\frac{\xi}{m}~.
\end{eqnarray}
Thus we solve the vacuum condition by
$\langle\CF_{\underline{\rm i}\underline{\rm i}}\rangle
=-2(\frac{e}{m}\pm i \frac{\xi}{m})$.
By fixing $SU(2)$ appropriately
we have managed to reproduce the conclusion of 
\cite{FIS1}. 

\section{$\CN=2$ $U(N_c)$ Gauge model with Hypermultiplets}

We generalize the $\CN=2$ $U(N_c)$ gauge model
to the $\CN=2$ $U(N_c)$ QCD model
coupled with the $q^+$- and $\omega$-hypermultiplets.
We consider both
the fundamental representation of $U(N_c)$
and the adjoint representation of $U(N_c)$.

\subsection{$q^+$ hypermultiplets}
The 
$q^+$ hypermultiplet
 is an analytic superfield
satisfying
$D^+q^{+}=\bar D^+q^{+}=0$,
and can be expanded as
\begin{eqnarray}
q^{+ }&=&
F^{+ }(x_A,u)
+\theta^{+}\psi (x_A,u)
+\bar\theta^+\bar\kappa (x_A,u)
+(\theta^+)^2M^{- }(x_A,u)
+(\bar\theta^+)^2N^{- }(x_A,u)
\nonumber\\&&
+i\theta^+\sigma^m\bar\theta^+A_m^{- }(x_A,u)
+(\theta^+)^2\bar\theta^+\bar\gamma^{(-2) }(x_A,u)
+(\bar\theta^+)^2\theta^+\chi^{(-2) }(x_A,u)
\nonumber\\&&
+(\theta^+)^2(\bar\theta^+)^2P^{(-3) }(x_A,u)~
\end{eqnarray}
in the analytic basis.
The physical fields are $SU(2)_A$ doublet complex scalars
(to be denoted as $f^i$)
contained in $F^+$
and a pair of $SU(2)_A$ isosinglet spinors,
$\psi$ and $\kappa$,
where $SU(2)_A$ is the automorphism of $\CN=2$.

We begin with a set of hypermultiplets $q^{+\mu}$
where
$\mu\equiv u =1,...,N_c$ for fundamental $q^+$
while $\mu\equiv a =0,1,...,N_c^2-1$ for adjoint $q^+$.
The $U(N_c)$ gauged matter action for $q^+$
is (see appendix C for the symplectic covariant form)
\begin{eqnarray}
S^{\rm gauged}_q&=&
-\intd ud\zeta^{(-4)}
\tilde q^+_{\mu}\CD^{++}q^{+\mu}~,~~~~
\CD^{++}q^{+\mu}=D^{++}q^{+\mu}+iV^{++a}(T_a)^\mu{}_\nu q^{+\nu}
\end{eqnarray}
where
$D^{++}=
\partial^{++}
-2i\theta^+\sigma^m\bar\theta^+\frac{\partial}{\partial x_A^m}
+\theta^{+\alpha}\frac{\partial}{\partial \theta^{-\alpha}}
+\bar\theta^{+\dot\alpha}\frac{\partial}{\partial \bar\theta^{-\dot\alpha}}
$
and
$\partial^{++}=u^{+i}\frac{\partial}{\partial u^{-i}}$.
$T_a$ is understood as
\begin{eqnarray}
(T_a)^\mu{}_\nu=
\left\{  
\begin{array}{ll}
(t_a)^u{}_v       & \mbox{for fundamental }q^+    \\
\ad(t_a)^b{}_c=if^b_{ac}       ~~~~~~~& \mbox{for adjoint }q^+      \\
  \end{array}
\right.
.
\end{eqnarray}
This action is invariant under the $U(N_c)$ gauge transformation
\begin{eqnarray}
\delta q^{+\mu}&=&i\epsilon^a
(T_a)^\mu{}_\nu q^{+\nu}~,~~~
\delta \tilde q^{+}_\mu=
-i\epsilon^a
\tilde q^{+}_\nu(T_a)^\nu{}_\mu~.
\end{eqnarray}
In other words, the $U(N_c)$ isometry
generated by the  Killing vector
$\lambda_a^{+\mu}\partial_{+\mu}
=i(T_a)^\mu{}_\nu q^{+\nu}\partial_{+\mu}$
has been gauged \cite{Bagger:1987rc}.
The Killing potential for $\lambda_a^{+\mu}$ is
given by
\begin{eqnarray}
\Lambda^{++}_{a}
=-\tilde q^{+}_\mu\lambda_a^{+\mu}
=-i\tilde q^{+}_\mu(T_a)^\mu{}_\nu q^{+\nu}~.
\end{eqnarray}

The equation of motion $\CD^{++}q^{+\mu}=0$
is expanded with respect to the order of $\theta$ as
\begin{eqnarray}
&&
\partial^{++}F^+=0~,
\label{eom 1}
\\&&
\partial^{++}\psi=\partial^{++}\bar\kappa=0~,
\label{eom 2}
\\&&
\partial^{++}M^-+\sqrt{2}\bar\phi F^+=0~,
\label{eom 3}
\\&&
\partial^{++}N^--\sqrt{2}\phi F^+=0~,
\label{eom 4}
\\&&
\partial^{++}A_m^--2\CD_mF^+=0~,
\label{eom 5}
\\&&
\partial^{++}\bar\gamma^{(-2)}
-i\bar\sigma^m\CD_m\psi
+\sqrt{2}\bar\phi\bar\kappa
-4i\bar\lambda^i F^+u_i^-=0~,
\label{eom 6}
\\&&
\partial^{++}\chi^{(-2)}
+i\sigma^m\CD_m\bar\kappa
-\sqrt{2}\phi\psi
+4i\lambda^i F^+u_i^-=0~,
\label{eom 7}
\\&&
\partial^{++}P^{(-3)}
-\CD^mA_m^-
+\sqrt{2}\bar\phi N^-
-\sqrt{2}\phi M^-
-2i\lambda^i\psi u_i^-
+2i\bar\lambda^i\bar\kappa u_i^-
+3iD^{ij}u_i^-u_j^-F^+=0~,
\nonumber\\&&
\label{eom 8}
\end{eqnarray}
where $\CD_m=\partial_m +iv_m$.
Here, we have omitted $U(N_c)$ index
understanding SW-NE contraction for the $U(N_c)$ index.
For example, $\phi F^+$ means 
$\phi^u{}_v F^{+v}=\phi^a(t_a)^u{}_vF^{+v}$ for fundamental $q^+$,
while $\phi^a{}_b F^{+b}=\phi^c\ad(t_c)^a{}_bF^{+b}=\phi^ci f^a_{cb}F^{+b}$ 
for adjoint $q^+$.
We find that these equations can be solved by
\begin{eqnarray}
(\ref{eom 1})&~\to ~&
F^+=f^i(x_A)u_i^+~,
\\
(\ref{eom 2})&~\to ~&
\psi=\psi(x_A)~,~~~
\bar\kappa=\bar\kappa(x_A)~,
\\
(\ref{eom 3})&~\to~&
M^-=-\sqrt{2}\bar\phi f^iu_i^-~,
\\
(\ref{eom 4})&~\to~&
N^-=\sqrt{2}\phi f^iu_i^-~,
\\
(\ref{eom 5})&~\to~&
A_m^-=2\CD_mf^iu_i^-~,
\\
(\ref{eom 6})&~\to~&
\bar\gamma^{(-2)}=2i\bar\lambda^if^ju_i^-u_j^-~,
\\&&
i\bar\sigma^m\CD_m\psi
-2i\bar\lambda^if_i
-\sqrt{2}\bar\phi\bar\kappa=0~,
\label{dynamical Psi}
\\
(\ref{eom 7})&~\to~&
\chi^{(-2)}=-2i\lambda^if^ju_i^-u_j^-~,
\\&&
i\sigma^m\CD_m\bar\kappa
-2i\lambda^if_i
-\sqrt{2}\phi\psi=0~,
\label{dynamical bar Psi}
\\
(\ref{eom 8})&~\to~&
P^{(-3)}=-iD^{ij}f^ku_{i}^-u_j^-u_{k}^-~,
\\&&
\CD^m\CD_mf^i
-(\phi\bar\phi+\bar\phi\phi)f^i
+i\lambda^i\psi
-i\bar\lambda^i\bar\kappa
+iD^{ij}f_j
=0~.
\label{dynamical f}
\end{eqnarray}
The infinitely many auxiliary fields contained in $q^{+}$
can be eliminated
by using these equations except for (\ref{dynamical Psi}),
(\ref{dynamical bar Psi}) and (\ref{dynamical f})
which are dynamical.
As a result,
the action $S^{\rm gauged}_{q}$ reduces to
\begin{eqnarray}
S_{q}^{\rm gauged}=
\intd^4x&\Big[&
-\bar f^i_\mu\CD^m\CD_m f_i^\mu
-\bar f_{i\mu}(\phi\bar\phi+\bar\phi\phi)^\mu{}_\nu f^{i\nu}
+i\bar f_{i\mu}D^{ij}{}^\mu{}_\nu f_j^\nu 
\nonumber\\&&
-\frac{i}{2}\bar\psi_\mu\bar\sigma^m\CD_m\psi^\mu
-\frac{i}{2}\kappa_\mu\sigma^m\CD_m\bar\kappa^\mu
+i\bar\psi_\mu\bar\lambda^i{}^\mu{}_\nu f_i^\nu 
-i\bar f^i_\mu\lambda_i{}^\mu{}_\nu \psi^\nu 
\nonumber\\&&
+i\kappa_\mu\lambda^i{}^\mu{}_\nu f_i^\nu 
+i\bar f^i_\mu\bar\lambda_i{}^\mu{}_\nu \bar\kappa^\nu 
+\frac{1}{\sqrt{2}}\kappa_\mu\phi^\mu{}_\nu \psi^\nu 
+\frac{1}{\sqrt{2}}\bar\psi_\mu\bar\phi^\mu{}_\nu \bar\kappa^\nu 
\Big]~.
\label{S:free compt}
\end{eqnarray}
The gauge transformation and the Killing potential reduce
respectively to
$\delta f^{\mu}_i=\epsilon^a (\ell_a)_{i}^{\mu}$ 
($\delta \bar f^{i}_\mu=\epsilon^a (\bar \ell_{a})_{\mu}^{i}$)
and $\Lambda^{++}_{a}|=\CQ^{ij}_au_i^+u_j^+$
where
\begin{eqnarray}
(\ell_{a})_i^\mu=i(T_a)^\mu{}_\nu f_i^{\nu}~,~~~
({\bar \ell_a})_{\mu}^{i}=i\bar f^{i}_\nu(T_a)^\nu{}_\mu~,~~~
\CQ^{ij}_a=\frac{i}{2}\left(
\bar f^i_\mu(T_a)^\mu{}_\nu f^{j\nu}+\bar f^j_\mu(T_a)^\mu{}_\nu f^{i\nu}
\right)~.~~~
\end{eqnarray}
Hence,
\begin{eqnarray}
-\bar f_i(\phi\bar\phi+\bar\phi\phi)f^i
=
(\bar\ell_a)_i(\ell_b)^i(\phi^a\bar\phi^b+\bar\phi^a\phi^b)~,~~~
i\bar f_iD^{ij}f_j=
\CQ^{ij}_aD^a_{ij}~.
\end{eqnarray}


We introduce flavors to the action $S^{\rm gauged}$.
We just regard the field $q^{+\mu}$ as an $N_f$-dimensional vector.
The action is simply
\begin{eqnarray}
S^{\rm gauged}_q&=&
-\intd ud\zeta^{(-4)}
\tilde q^+_{\mu I}\CD^{++}q^{+\mu I}~,
\end{eqnarray}
where $I=1,...,N_f$.
This implies that component fields
$f$, $\psi$ and $\kappa$ are understood as $N_f$-dimensional vectors.
We suppress the flavor index in the followings.

In summary,
the $\CN=2$ $U(N_c)$ gauged action for $N_f$ hypermultiplets $q^{+u}$ 
in the fundamental representation of $U(N_c)$ 
is
\begin{eqnarray}
S_{{\rm fund}\,q}^{\rm gauged}=
\intd^4x&\Bigg[&
-\bar f^i_u\CD^m\CD_m f_i^u
-\frac{i}{2}\bar\psi_u\bar\sigma^m\CD_m\psi^u
-\frac{i}{2}\kappa_u\sigma^m\CD_m\bar\kappa^u
\nonumber\\&&
-\bar f_{iu}(\phi\bar\phi+\bar\phi\phi)^u{}_vf^{iv}
+\hat\CQ_{a}^{ij}D_{ij}^a
\nonumber\\&&
+\Big(
-i\bar f^i_u\lambda_i{}^u{}_v\psi^v
+i\kappa_u\lambda^i{}^u{}_vf_i^v
+\frac{1}{\sqrt{2}}\kappa_u\phi^u{}_v\psi^v
+c.c.
\Big)
\Bigg]~
\label{S:free fund compt}
\end{eqnarray}
where $\hat\CQ^{ij}_a\equiv\CQ^{ij}_a|_{T_a=t_a}$.
On the other hand, the $\CN=2$ $U(N_c)$ gauged action for 
$N_a$ hypermultiplets $q^{+a}$ 
in the adjoint representation of $U(N_c)$ 
is
\begin{eqnarray}
S_{{\rm adj}\,q}^{\rm gauged}=
\intd^4x&\Bigg[&
-\bar f^i_a\CD^m\CD_m f_i^a
-\frac{i}{2}\bar\psi_a\bar\sigma^m\CD_m\psi^a
-\frac{i}{2}\kappa_a\sigma^m\CD_m\bar\kappa^a
\nonumber\\&&
-\bar f_{ia}(\phi\bar\phi+\bar\phi\phi)^a{}_bf^{ib}
+\check\CQ_{a}^{ij}D_{ij}^a
\nonumber\\&&
+\Big(
-i\bar f^i_a\lambda_i{}^a{}_b\psi^b
+i\kappa_a\lambda^i{}^a{}_bf_i^b
+\frac{1}{\sqrt{2}}\kappa_a\phi^a{}_b\psi^b
+c.c.
\Big)
\Bigg]~
\label{S:free adj compt}
\end{eqnarray}
where $\check\CQ^{ij}_a\equiv\CQ^{ij}_a|_{T_a=\ad(t_a)}$.

\subsection{$\omega$ hypermultiplets}

The $\omega$ hypermultiplet is known as a real hypermultiplet, $\omega=\widetilde{\omega}$.
Here we combine two of them
to a complex superfield,
or equivalently we do not impose the reality condition.
Such a complex $\boldsymbol{\omega}$ hypermultiplet is expanded as
\begin{eqnarray}
\boldsymbol{\omega}(\zeta,u)&=&
\omega(x_A,u)
+\theta^+\psi^-(x_A,u)
+\bar\theta^+\bar\kappa^-(x_A,u)
+(\theta^+)^2M^{--}(x_A,u)
+(\bar\theta^+)^2N^{--}(x_A,u)
\nonumber\\&&
+i\theta^+\sigma^m\bar\theta^+A_m^{--}(x_A,u)
+(\theta^+)^2\bar\theta^+\bar\gamma^{(-3)}(x_A,u)
+(\bar\theta^+)^2\theta^+\chi^{(-3)}(x_A,u)
\nonumber\\&&
+(\theta^+)^4P^{(-4)}(x_A,u)~.
\label{omega:component}
\end{eqnarray}

We consider the $U(N_c)$ gauged action
\begin{eqnarray}
S^{\rm gauged}_\omega&=&
\frac{1}{2}\intd ud\zeta^{(-4)}
\widetilde{\boldsymbol{\omega}}(\CD^{++})^2\boldsymbol{\omega}~,~~~
\CD^{++}=D^{++}+iV^{++}~,~~~V^{++}=V^{++a}T_a~,~~
\end{eqnarray}
where $T_a=(t_a)^u{}_v$, $u,v=1,\cdots,N_c$
for $\boldsymbol{\omega}^u$ in the fundamental representation of $U(N_c)$ ,
while $T_a=\ad(t_a)^b{}_c=if^b_{ac}$, $a,b=0,1,\cdots,N_c^2-1$
for $\boldsymbol{\omega}^a$ in the adjoint representation of $U(N_c)$ .
The equation of motion,
$(\CD^{++})^2\boldsymbol{\omega}=0$, 
is expanded 
with respect to the order of $\theta$
as
\begin{eqnarray}
&&(\partial^{++})^2\omega=0~,
\label{w:eom 1}\\
&&(\partial^{++})^2\psi^-=
(\partial^{++})^2\bar\kappa^-=
0~,
\label{w:eom 2}\\
&&(\partial^{++})^2M^{--}
+2\sqrt{2}\bar\phi\partial^{++}\omega=0~,
\label{w:eom 3}\\
&&(\partial^{++})^2N^{--}
-2\sqrt{2}\phi\partial^{++}\omega=0~,
\label{w:eom 4}\\
&&(\partial^{++})^2A_m^{--}
-4\partial^{++}\CD_m\omega=0~,
\label{w:eom 5}\\
&&(\partial^{++})^2\bar\gamma^{(-3)}
-2i\bar\sigma^m\CD_m\partial^{++}\psi^-
+2\sqrt{2}\bar\phi\partial^{++}\bar\kappa^-
-8i\bar\lambda^iu_i^-\partial^{++}\omega
-4i\bar\lambda^iu_i^+\omega
=0~,
\label{w:eom 6}\\
&&(\partial^{++})^2\chi^{(-3)}
+2i\sigma^m\CD_m\partial^{++}\bar\kappa^-
-2\sqrt{2}\phi\partial^{++}\psi^-
+8i\lambda^iu_i^-\partial^{++}\omega
+4i\lambda^iu_i^+\omega
=0~,
\label{w:eom 7}\\
&&(\partial^{++})^2P^{(-4)}
-2\CD^m\partial^{++}A_m^{--}
+2\CD^m\CD_m\omega
+2\sqrt{2}\bar\phi\partial^{++}N^{--}
-2\sqrt{2}\phi\partial^{++}M^{--}
\nonumber\\&&~~~
-2i\lambda^iu_i^+\psi^-
-4i\lambda^iu_i^-\partial^{++}\psi^-
+2i\bar\lambda^iu_i^+\bar\kappa^-
+4i\bar\lambda^iu_i^-\partial^{++}\bar\kappa^-
\nonumber\\&&~~~
+6iD^{ij}u_i^+u_j^-\omega
+6iD^{ij}u_i^-u_j^-\partial^{++}\omega
-2(\bar\phi\phi+\phi\bar\phi)\omega
=0~.
\label{w:eom 8}
\end{eqnarray}
We find that these are solved by
\begin{eqnarray}
(\ref{w:eom 1})&~\to~&
\omega(x_A,u)=\frac{1}{\sqrt{2}}\omega(x_A)+\omega^{ij}(x_A)u_{(i}^+u_{j)}^-~,
\label{w:sol 1}\\
(\ref{w:eom 2})&~\to~&
\psi^-(x_A,u)=\psi^i(x_A)u_i^-~,~~~
\bar\kappa^-(x_A,u)=\bar\kappa^i(x_A)u_i^-~,
\label{w:sol 2}\\
(\ref{w:eom 3})&~\to~&
M^{--}(x_A,u)=
-\sqrt{2}\bar\phi\omega^{ij}u_i^-u_j^-~,
\label{w:sol 3}\\
(\ref{w:eom 4})&~\to~&
N^{--}(x_A,u)=
\sqrt{2}\phi\omega^{ij}u_i^-u_j^-~,
\label{w:sol 4}\\
(\ref{w:eom 5})&~\to~&
A_m^{--}(x_A,u)=
2\CD_m\omega^{ij}u_i^-u_j^-~,
\label{w:sol 5}\\
(\ref{w:eom 6})&~\to~&
\bar\gamma^{(-3)}(x_A,u)=
2i\bar\lambda^i\omega^{jk}u_i^-u_j^-u_k^-~,
\label{w:sol 6}\\&&
i\bar\sigma^m\CD_m\psi^i
-\sqrt{2}\bar\phi\bar\kappa^i
+2i\bar\lambda_j\omega^{ji}
+\sqrt{2}i\bar\lambda^i\omega
=0~,
\label{w:sol 6 dynamical}\\
(\ref{w:eom 7})&~\to~&
\chi^{(-3)}(x_A,u)=
-2i\lambda^i\omega^{jk}u_i^-u_j^-u_k^-~,
\label{w:sol 7}\\&&
i\sigma^m\CD_m\bar\kappa^i
-\sqrt{2}\phi\psi^i
+2i\lambda_j\omega^{ji}
+\sqrt{2}i\lambda^i\omega
=0~,
\label{w:sol 7 dynamical}\\
(\ref{w:eom 8})&~\to~&
P^{(-4)}(x_A,u)=
-iD^{ij}\omega^{kl}u_i^-u_j^-u_k^-u_l^-~,
\label{w:sol 8}
\\&&
\CD^m\CD_m\omega^{ij}
-(\bar\phi\phi+\phi\bar\phi)\omega^{ij}
-\frac{i}{\sqrt{2}}D^{ij}\omega
-iD^i{}_k\omega^{kj}
+i\lambda^{(i}\psi^{j)}
-i\bar\lambda^{(i}\bar\kappa^{j)}
=0~,\nonumber\\
\label{w:sol 8 dynamical 1}
\\&&
\sqrt{2}\CD^m\CD_m\omega
+iD^{ij}\omega_{ij}
+i\lambda^i\psi_i
-i\bar\lambda^i\bar\kappa_i
-\sqrt{2}(\bar\phi\phi+\phi\bar\phi)\omega=0
~.
\label{w:sol 8 dynamical 2}
\end{eqnarray}
Eliminating infinitely many auxiliary fields
by using these equations except for
(\ref{w:sol 6 dynamical}),
(\ref{w:sol 7 dynamical}),
(\ref{w:sol 8 dynamical 1})
and (\ref{w:sol 8 dynamical 2})
which are dynamical,
the action reduces to
\begin{eqnarray}
S^{\rm gauged}_{\omega}=\intd x^4&\Big[&
+\frac{1}{2}\bar\omega^{ij}\CD^m\CD_m\omega_{ij}
+\frac{1}{2}\bar\omega\CD^m\CD_m\omega
-\frac{i}{4}\kappa^i\sigma^m\CD_m\bar\kappa_i
+\frac{i}{4}\bar\psi^i\bar\sigma^m\CD_m\psi_i
\nonumber\\&&
-\frac{1}{2}\bar\omega^{ij}(\bar\phi\phi+\phi\bar\phi)\omega_{ij}
-\frac{1}{2}\bar\omega(\bar\phi\phi+\phi\bar\phi)\omega
+\CS^{ij}_aD^a_{ij}
\nonumber\\&&
+\frac{\sqrt{2}}{4}\kappa^i\phi\psi_i
+\frac{i}{2}\kappa^i\lambda^j\omega_{ij}
-\frac{\sqrt{2}}{4}i\kappa^i\lambda_i\omega
+\frac{i}{2}\bar\omega^{ij}\lambda_i\psi_j
+\frac{\sqrt{2}}{4}i\bar\omega\lambda^i\psi_i
\nonumber\\&&
-\frac{\sqrt{2}}{4}\bar\psi^i\bar\phi\bar\kappa_i
-\frac{i}{2}\bar\omega^{ij}\bar\lambda_i\bar\kappa_j
-\frac{\sqrt{2}}{4}i\bar\omega\bar\lambda^i\bar\kappa_i
-\frac{i}{2}\bar\psi^i\bar\lambda^j\omega_{ij}
+\frac{\sqrt{2}}{4}i\bar\psi^i\bar\lambda_i\omega
~\Big]~~~
\label{action w: component}
\end{eqnarray}
where
\begin{eqnarray}
\CS^{ij}_a\equiv
\frac{i}{2}\bar\omega^{ki}T_a\omega^j{}_k
+\frac{\sqrt{2}}{4}i\bar\omega^{ij}T_a\omega
-\frac{\sqrt{2}}{4}i\bar\omega T_a\omega^{ij}~.
\end{eqnarray}
It is easy to introduce flavors to the action by regarding $\boldsymbol{\omega}$
as a vector $\boldsymbol{\omega}^I$
\begin{eqnarray}
S^{\rm gauged}_\omega&=&
\frac{1}{2}\intd ud\zeta^{(-4)}
\widetilde{\boldsymbol{\omega}}_I(\CD^{++})^2\boldsymbol{\omega}^I~,
\end{eqnarray}
where $I=1,\cdots,N_f^\omega$ for fundamental $\omega$
and $I=1,\cdots,N_a^\omega$ for adjoint $\omega$.
We omit the flavor index below.

In summary, the $U(N_c)$ gauged action for
$N_f^\omega$ hypermultiplets $\omega^u$ in the fundamental representation
of $U(N_c)$
is
\begin{eqnarray}
&&S^{\rm gauged}_{{\rm fund}\,\omega}=
\nonumber\\&&
\intd x^4\Bigg[
\frac{1}{2}\bar\omega^{ij}_u\CD^m\CD_m\omega_{ij}^u
+\frac{1}{2}\bar\omega_u\CD^m\CD_m\omega^u
-\frac{i}{4}\kappa^i_u\sigma^m\CD_m\bar\kappa_i^u
+\frac{i}{4}\bar\psi^i_u\bar\sigma^m\CD_m\psi_i^u
\nonumber\\&&
-\frac{1}{2}\bar\omega^{ij}_u(\bar\phi\phi+\phi\bar\phi)^u{}_v\omega_{ij}^v
-\frac{1}{2}\bar\omega_u(\bar\phi\phi+\phi\bar\phi)^u{}_v\omega^v
+\hat\CS^{ij}_aD^a_{ij}
\label{action w: component:fund}
\\&&
+\Big(
\frac{\sqrt{2}}{4}\kappa^i_u\phi^u{}_v\psi_i^v
+\frac{i}{2}\kappa^i_u\lambda^{ju}{}_v\omega_{ij}^v
-\frac{\sqrt{2}}{4}i\kappa^i_u\lambda_i{}^u{}_v\omega^v
+\frac{i}{2}\bar\omega^{ij}_u\lambda_i^u{}_v\psi_j^v
+\frac{\sqrt{2}}{4}i\bar\omega_u\lambda^{iu}{}_v\psi_i^v
+c.c.
\Big)
~\Bigg]
\nonumber
\end{eqnarray}
with $\hat \CS^{ij}_a=\CS^{ij}_a|_{T_a=t_a}$,
while that for
$N_a^\omega$ hypermultiplets $\omega^a$ in the adjoint representation
of $U(N_c)$
is
\begin{eqnarray}
&&S^{\rm gauged}_{{\rm adj}\,\omega}=
\nonumber\\&&
\intd x^4\Bigg[
\frac{1}{2}\bar\omega^{ij}_a\CD^m\CD_m\omega_{ij}^a
+\frac{1}{2}\bar\omega_a\CD^m\CD_m\omega^a
-\frac{i}{4}\kappa^i_a\sigma^m\CD_m\bar\kappa_i^a
+\frac{i}{4}\bar\psi^i_a\bar\sigma^m\CD_m\psi_i^a
\nonumber\\&&
-\frac{1}{2}\bar\omega^{ij}_a(\bar\phi\phi+\phi\bar\phi)^a{}_b\omega_{ij}^b
-\frac{1}{2}\bar\omega_a(\bar\phi\phi+\phi\bar\phi)^a{}_b\omega^b
+\check\CS^{ij}_aD^a_{ij}
\label{action w: component:adj}
\\&&
+\Big(
\frac{\sqrt{2}}{4}\kappa^i_a\phi^a{}_b\psi_i^b
+\frac{i}{2}\kappa^i_a\lambda^{ja}{}_b\omega_{ij}^b
-\frac{\sqrt{2}}{4}i\kappa^i_a\lambda_i{}^a{}_b\omega^b
+\frac{i}{2}\bar\omega^{ij}_a\lambda_i^a{}_b\psi_j^b
+\frac{\sqrt{2}}{4}i\bar\omega_a\lambda^{ia}{}_b\psi_i^b
+c.c.
\Big)
~\Bigg]
\nonumber
\end{eqnarray}
with $\check \CS^{ij}_a=\CS^{ij}_a|_{T_a=\ad(t_a)}$.

\medskip

The $\CN=2$ $U(N_c)$  QCD action
coupled with $N_f$ fundamental $q^{+u}$,
$N_a$ adjoint $q^{+a}$,
$N_f^\omega$ fundamental $\omega^u$
and
$N_a^\omega$ adjoint $\omega^a$
is given as 
\begin{eqnarray}
S_{V,q,\omega}=
S_V
+S_{{\rm fund}\,q}^{\rm gauged}
+S_{{\rm adj}\,q}^{\rm gauged}
+S_{{\rm fund}\,\omega}^{\rm gauged}
+S_{{\rm adj}\,\omega}^{\rm gauged}~
\label{action: matter}
\end{eqnarray}
with (\ref{SV:component}),
(\ref{S:free fund compt}),
(\ref{S:free adj compt}),
(\ref{action w: component:fund}) and (\ref{action w: component:adj}).

\subsection{electric \& magnetic FI terms}
Next, we add the electric and magnetic FI terms to the action
$S_{V,q,\omega}$ in (\ref{action: matter}).
The electric FI term $S_e$ is the same as (\ref{Se}).
As was seen in subsection 2.1, the effect of the magnetic FI term
is to shift the auxiliary field $D$
by an imaginary constant.
Thanks to this property the magnetic FI term
does not affect the superinvariance of the total action
which now includes $S_e$.
This observation leads to the magnetic FI term of the form
\begin{eqnarray}
S^{\rm QCD}_m&=&S_m^{\rm YM}
+\intd^4x~
2i(\hat\CQ_0^{ij}+\hat\CS_0^{ij})(\xi^{ij}_D-\bar\xi^{ij}_D)
\label{mFI:QCD}
\end{eqnarray}
where $S_m^{\rm YM}$ is given in (\ref{mag.FI:YM}).
The last term in  (\ref{mFI:QCD})
shifts the terms, 
$\hat\CQ_{0}^{ij}D_{ij}^0$ in $S^{\rm gauged}_{{\rm fund}\,q}$
(\ref{S:free fund compt})
and $\hat\CS_{0}^{ij}D_{ij}^0$ in $S^{\rm gauged}_{{\rm fund}\,\omega}$
(\ref{action w: component:fund}),
as
\begin{eqnarray}
\hat\CQ_{a}^{ij}D_{ij}^a+\hat\CS_{a}^{ij}D_{ij}^a
+ 2i(\hat\CQ_0^{ij}+\hat\CS_0^{ij})(\xi^{ij}_D-\bar\xi^{ij}_D)
&=&
\frac{1}{2}(\hat\CQ_a^{ij}+\hat\CS_a^{ij})
\boldsymbol{D}_{ij}^a+c.c.
~
\end{eqnarray}
where $\boldsymbol{D}$ is given in (\ref{bold D}).
The terms, 
$\check\CQ_{0}^{ij}D_{ij}^0$ in $S^{\rm gauged}_{{\rm adj}\,q}$
(\ref{S:free adj compt})
and
$\check\CS_{0}^{ij}D_{ij}^0$ in $S^{\rm gauged}_{{\rm adj}\,\omega}$
(\ref{action w: component:adj}),
vanish because
$\ad(t_0)=0$,
and so the adjoint matters do not appear in (\ref{mFI:QCD}). 
Thus, we observe that
\begin{eqnarray}
S_{V,q,\omega}+S_m^{\rm QCD}=S_{V,q,\omega}|_{D\to\boldsymbol{D}}
\end{eqnarray}
where $|_{D\to\boldsymbol{D}}$ means 
the replacement $D^{aij}\to \boldsymbol{D}^{aij}$
($D^{aij}\to \bar{\boldsymbol{D}}^{aij}$).

\bigskip

Gathering all together,
the total action is 
\begin{eqnarray}
S_{\rm QCD}=S_{V,q,\omega}
+S_e+S_m^{\rm QCD}~,
\label{S_QCD}
\end{eqnarray}
with (\ref{action: matter}),
(\ref{Se}) and (\ref{mFI:QCD}).
The auxiliary field $D^{aij}$ is solved by
\begin{eqnarray}
D^{aij}=
-2g^{ab}&\Big[&
(\xi^{ij}+\bar\xi^{ij})\delta_b^0
+\xi_D^{ij}\CF_{0b}|+\bar\xi^{ij}_D\overline{\CF}_{0b}|
+\hat\CQ^{ij}_{b}
+\check\CQ^{ij}_{b}
+\hat\CS^{ij}_{b}
+\check\CS^{ij}_{b}
\nonumber\\&&
+\frac{i}{4}\CF_{bcd}|\lambda^{ci}\lambda^{dj}
-\frac{i}{4}\overline{\CF}_{bcd}|\bar\lambda^{ci}\bar\lambda^{dj}
\Big]~
\label{D:QCD}
\end{eqnarray}
while the $\boldsymbol{D}^{aij}$ is given as
\begin{eqnarray}
\boldsymbol{D}^{aij}=
-2g^{ab}&\Big[&
(\xi^{ij}+\bar\xi^{ij})\delta_b^0
+(\xi_D^{ij}+\bar\xi^{ij}_D)\overline{\CF}_{0b}|
+\hat\CQ^{ij}_b
+\check\CQ^{ij}_b
+\hat\CS^{ij}_{b}
+\check\CS^{ij}_{b}
\nonumber\\&&
+\frac{i}{4}\CF_{bcd}|\lambda^{ci}\lambda^{dj}
-\frac{i}{4}\overline{\CF}_{bcd}|\bar\lambda^{ci}\bar\lambda^{dj}
\Big]~.
\label{bold D:QCD}
\end{eqnarray}
After eliminating auxiliary field $D$ by using (\ref{D:QCD}),
the action of the $\CN=2$ $U(N_c)$ QCD model becomes
\begin{eqnarray}
S_{\rm QCD}'&=&\intd^4x\Big[
\CL_{\rm kin}+\CL_{\rm pot}+\CL_{\rm Pauli}+\CL_{\rm mass}+\CL_{\rm 4\,fermi}
\Big]
\label{S'_QCD}
\end{eqnarray}
where
\begin{eqnarray}
\CL_{\rm kin}&=&
-g_{ab}\CD^m\phi^a \CD_m\bar\phi^b
-\frac{1}{2}\overline{\CF}_{ab}|\bar\lambda^{ai}\bar\sigma^m\CD_m\lambda^b_i
-\frac{1}{2}{\CF}_{ab}|\lambda^{ai}\sigma^m\CD_m\bar\lambda^{b}_i
\nonumber\\&&
-\frac{1}{4}g_{ab}v_{mn}^av^{b\,mn}
-\frac{1}{8}\real\CF_{ab}|\varepsilon^{mnpq}v^a_{mn}v^b_{pq}
\nonumber\\&&
-\bar f^i_u\CD^m\CD_m f_i^u
-\frac{i}{2}\bar\psi_u\bar\sigma^m\CD_m\psi^u
-\frac{i}{2}\kappa_u\sigma^m\CD_m\bar\kappa^u
\nonumber\\&&
-\bar f^i_a\CD^m\CD_m f_i^a
-\frac{i}{2}\bar\psi_a\bar\sigma^m\CD_m\psi^a
-\frac{i}{2}\kappa_a\sigma^m\CD_m\bar\kappa^a
\nonumber\\&&
+\frac{1}{2}\bar\omega^{ij}_u\CD^m\CD_m\omega_{ij}^u
+\frac{1}{2}\bar\omega_u\CD^m\CD_m\omega^u
-\frac{i}{4}\kappa^i_u\sigma^m\CD_m\bar\kappa_i^u
+\frac{i}{4}\bar\psi^i_u\bar\sigma^m\CD_m\psi_i^u
\nonumber\\&&
+
\frac{1}{2}\bar\omega^{ij}_a\CD^m\CD_m\omega_{ij}^a
+\frac{1}{2}\bar\omega_a\CD^m\CD_m\omega^a
-\frac{i}{4}\kappa^i_a\sigma^m\CD_m\bar\kappa_i^a
+\frac{i}{4}\bar\psi^i_a\bar\sigma^m\CD_m\psi_i^a
~,\\
\CL_{\rm pot}&=&
-\frac{1}{4}g_{ab}D^{aij}|~ D^{b}_{ij}|
-g_{ab}\CP^a\CP^b
+2i\xi^{ij}_D\xi_{Dij}\CF_{00}|
-2i\bar\xi^{ij}_D\bar\xi_{Dij}\overline{\CF}_{00}|
\nonumber\\&&
+2i(\hat\CQ_0+\hat\CS_0)(\xi^{ij}_D-\bar\xi^{ij}_D)
-2i(\xi^{ij}+\bar\xi^{ij})(\xi_{Dij}-\bar\xi_{Dij})
\nonumber\\&&
-\bar f^i_u(\bar\phi\phi+\phi\bar\phi)^u{}_vf_i^v
-\frac{1}{2}\bar\omega^{ij}_u(\bar\phi\phi+\phi\bar\phi)^u{}_v\omega_{ij}^v
-\frac{1}{2}\bar\omega_u(\bar\phi\phi+\phi\bar\phi)^u{}_v\omega^v
\nonumber\\&&
-\bar f^i_a(\bar\phi\phi+\phi\bar\phi)^a{}_bf_i^b
-\frac{1}{2}\bar\omega^{ij}_a(\bar\phi\phi+\phi\bar\phi)^a{}_b\omega_{ij}^b
-\frac{1}{2}\bar\omega_a(\bar\phi\phi+\phi\bar\phi)^a{}_b\omega^b
\nonumber\\
&=&
-\frac{1}{4}g_{ab}\boldsymbol{D}^{aij}|~ \bar{\boldsymbol{D}}^{b}_{ij}|
-g_{ab}\CP^a\CP^b
-2i(\xi^{ij}+\bar\xi^{ij})(\xi_{Dij}-\bar\xi_{Dij})
\nonumber\\&&
-\bar f^i_u(\bar\phi\phi+\phi\bar\phi)^u{}_vf_i^v
-\frac{1}{2}\bar\omega^{ij}_u(\bar\phi\phi+\phi\bar\phi)^u{}_v\omega_{ij}^v
-\frac{1}{2}\bar\omega_u(\bar\phi\phi+\phi\bar\phi)^u{}_v\omega^v
\nonumber\\&&
-\bar f^i_a(\bar\phi\phi+\phi\bar\phi)^a{}_bf_i^b
-\frac{1}{2}\bar\omega^{ij}_a(\bar\phi\phi+\phi\bar\phi)^a{}_b\omega_{ij}^b
-\frac{1}{2}\bar\omega_a(\bar\phi\phi+\phi\bar\phi)^a{}_b\omega^b
~,
\label{Lpot:QCD}
\\
\CL_{\rm mass}&=&
+\frac{1}{2}g_{ab}\lambda^{ai}f^b_{cd}(-i\sqrt{2}\bar\phi^c)\lambda^d_i
+\frac{i}{4}\boldsymbol{D}^{aij}|~\CF_{abc}|\lambda^b_i\lambda^c_j
\nonumber\\&&
-i\bar f^i_u\lambda_i{}^u{}_v\psi^v
+i\kappa_u\lambda^i{}^u{}_vf_i^v
+\frac{1}{\sqrt{2}}\kappa_u\phi^u{}_v\psi^v
-i\bar f^i_a\lambda_i{}^a{}_b\psi^b
+i\kappa_a\lambda^i{}^a{}_bf_i^b
+\frac{1}{\sqrt{2}}\kappa_a\phi^a{}_b\psi^b
\nonumber\\&&
+
\frac{\sqrt{2}}{4}\kappa^i_u\phi^u{}_v\psi_i^v
+\frac{i}{2}\kappa^i_u\lambda^{ju}{}_v\omega_{ij}^v
-\frac{\sqrt{2}}{4}i\kappa^i_u\lambda_i{}^u{}_v\omega^v
+\frac{i}{2}\bar\omega^{ij}_u\lambda_i^u{}_v\psi_j^v
+\frac{\sqrt{2}}{4}i\bar\omega_u\lambda^{iu}{}_v\psi_i^v
\nonumber\\&&
+
\frac{\sqrt{2}}{4}\kappa^i_a\phi^a{}_b\psi_i^b
+\frac{i}{2}\kappa^i_a\lambda^{ja}{}_b\omega_{ij}^b
-\frac{\sqrt{2}}{4}i\kappa^i_a\lambda_i{}^a{}_b\omega^b
+\frac{i}{2}\bar\omega^{ij}_a\lambda_i^a{}_b\psi_j^b
+\frac{\sqrt{2}}{4}i\bar\omega_a\lambda^{ia}{}_b\psi_i^b
\nonumber\\&&
+c.c.
~,
\label{Lmass:QCD}
\end{eqnarray}
where
\begin{eqnarray}
D^{aij}|&=&-2g^{ab}\Big[
(\xi^{ij}+\bar\xi^{ij})\delta_b^0
+\xi^{ij}_D\CF_{0b}|+\bar\xi^{ij}_D\overline{\CF}_{0b}|
+\hat\CQ_b^{ij}
+\check\CQ_b^{ij}
+\hat\CS^{ij}_{b}
+\check\CS^{ij}_{b}
\Big]~,\\
\boldsymbol{D}^{aij}|&=&D^{aij}|+4i\xi^{ij}_D\delta^a_0
\nonumber\\
&=&
-2g^{ab}\Big[
(\xi^{ij}+\bar\xi^{ij})\delta_b^0
+(\xi^{ij}_D+\bar\xi^{ij}_D)\overline{\CF}_{0b}|
+\hat\CQ_b^{ij}
+\check\CQ_b^{ij}
+\hat\CS^{ij}_{b}
+\check\CS^{ij}_{b}
\Big]~.
\end{eqnarray}
$\CL_{\rm Pauli}$,
$\CL_{\rm 4\,fermi}$
and $\CP^a$
are
given in (\ref{YM:pauli}), (\ref{YM:4 fermi})
and (\ref{P}), respectively.
The action $S_{\rm QCD}'$ without FI terms 
corresponds to
the $\CN=2$ action
(see for example \cite{N=2 sugra})
with flat four-dimensional space-time
and with flat hyperk\"ahler geometry of hypermultiplets.

\section{Vacua of $\CN=2$ $U(N_c)$ QCD model}
Let us examine vacua of the $\CN=2$ $U(N_c)$ QCD model
coupled with hypermultiplets 
and equipped with
the electric and magnetic FI terms,
$S_{\rm QCD}'$ in (\ref{S'_QCD}).

The vacua are determined by the scalar potential
\begin{eqnarray}
\CV&=&
\frac{1}{4}g_{ab}\boldsymbol{D}^{aij}|\, \bar{\boldsymbol{D}}^{b}_{ij}|
+g_{ab}\CP^a\CP^b
+2i(\xi^{ij}+\bar\xi^{ij})(\xi_{Dij}-\bar\xi_{Dij})
\nonumber\\&&
+\overline{f^i}_u(\bar\phi\phi+\phi\bar\phi)^u{}_vf^{iv}
+\frac{1}{2}\bar\omega^{ij}_u(\bar\phi\phi+\phi\bar\phi)^u{}_v\omega_{ij}^v
+\frac{1}{2}\bar\omega_u(\bar\phi\phi+\phi\bar\phi)^u{}_v\omega^v
\nonumber\\&&
+\overline{f^i}_a(\bar\phi\phi+\phi\bar\phi)^a{}_bf^{ib}
+\frac{1}{2}\bar\omega^{ij}_a(\bar\phi\phi+\phi\bar\phi)^a{}_b\omega_{ij}^b
+\frac{1}{2}\bar\omega_a(\bar\phi\phi+\phi\bar\phi)^a{}_b\omega^b
~.
\end{eqnarray}
We demand positive definiteness  of $g_{ab}$. 
The second term in the first line tells us that $\langle\phi^r\rangle=0$ where 
$r \in$ non-Cartan directions.
The first term in the second line reduces to
\begin{eqnarray}
\langle
|\phi^{\underline{\rm i}}|^2\overline{f^i}_{\underline{\rm i}}
f^{i{\underline{\rm i}}}
\rangle
\end{eqnarray}
because $\langle\phi\bar\phi\rangle=
\langle|\phi^{\underline{\rm i}}|^2\rangle t_{\underline{\rm i}}$.
Let us examine the phase in which $\langle\phi^{\underline{\rm i}}\rangle\neq 0$
(so called the Coulomb phase).
This leads to $\langle f^i_{\underline{\rm i}}\rangle=0$,
 i.e.,
 $\langle f^i_u\rangle=0$,
 which implies that $\langle\hat\CQ_b^{ij}\rangle=0$.
Similarly one finds that 
$\langle\omega_{ij}^u\rangle=\langle\omega^u\rangle=0$
and $\langle\hat\CS_b^{ij}\rangle=0$.
On the other hand,
the first term in the last line becomes
\begin{eqnarray}
\langle
\overline{f^i}_r\phi^{\underline{\rm i}}\bar\phi^{\underline{\rm j}}
(f^r_{{\underline{\rm i}}s}f^s_{{\underline{\rm j}}t}
+f^r_{{\underline{\rm j}}s}f^s_{{\underline{\rm i}}t}
)f^t_i
\rangle
\end{eqnarray}
and thus we derive $\langle f^i_r\rangle=0$
where $r \in$ non-Cartan directions,
which implies that $\langle\check\CQ^{ij}_b\rangle=0$.
Similarly we derive 
$\langle\omega_{ij}^r\rangle=\langle\omega^r\rangle=0$
and $\langle\check\CS_b^{ij}\rangle=0$.
Summarizing we derived vevs at the vacua
\begin{eqnarray}
\langle f^i_u\rangle=\langle \omega^{ij}_u\rangle=\langle\omega_u\rangle=0~~~
\mbox{and}~~~
\langle f^i_r\rangle=\langle \omega^{ij}_r\rangle=\langle\omega_r\rangle=0~.
\label{vev:matter}
\end{eqnarray}
Finally we examine the first term in the first line
in $\CV$, say $\CV_1$.
By using (\ref{vev:matter}),
we derive
\begin{eqnarray}
\langle\partial_{A}\CV_1\rangle=0~,~~~
A=\{f^i_u,\omega^{ij}_u,\omega_u,
f^i_a,\omega^{ij}_a,\omega_a,\}
~.
\end{eqnarray}
We note here that 
$\langle f^i_{\underline{{\rm i}}}\rangle$,
$\langle \omega^{ij}_{\underline{{\rm i}}}\rangle$ and
$\langle\omega_{\underline{{\rm i}}}\rangle$
are not determined by the scalar potential $\CV$.
In addition, $\partial_{\bar A}\partial_A\CV=0$ for
$A=\{f^i_{\underline{{\rm i}}},
\omega^{ij}_{\underline{{\rm i}}},
\omega_{\underline{{\rm i}}}
\}$.
This means that the adjoint matter scalars in the Cartan direction
parametrize the flat directions in $\CV$.
The non-trivial vacuum condition is
\begin{eqnarray}
\langle\partial_{\phi^a}\CV_1\rangle=
\frac{i}{8}\langle\CF_{abc}|~\boldsymbol{D}^{bij}\boldsymbol{D}^c_{ij}\rangle
=\frac{i}{4}\langle\CF_{abc}|~\boldsymbol{D}^{bA}\boldsymbol{D}^{cA}\rangle
\end{eqnarray}
where $\langle\boldsymbol{D}^{aA}\rangle$
is the same as one without hypermultiplets in (\ref{bold D:YM})
because 
$\langle\hat\CQ^{ij}_a\rangle=
\langle\check\CQ^{ij}_a\rangle=
\langle\hat\CS^{ij}_a\rangle=
\langle\check\CS^{ij}_a\rangle=0$.
In this way,
we arrive at the same vacuum condition
as one for the model specified by $S_{\rm YM}'$ in (\ref{action:YM'}).
As was explained in section 3,
the supersymmetry transformation of $\lambda^i$ on the vacua is
\begin{eqnarray}
\langle\delta_\eta(U^1{}_j\lambda^{\underline{\rm i}j})\rangle\neq 0~,~~~
\langle\delta_\eta(U^2{}_j\lambda^{\underline{\rm i}j})\rangle= 0~,~~~
\langle\delta_\eta\lambda^{ri}\rangle= 0~.
\end{eqnarray}
The supersymmetry transformation of matter fermions in $q^+$ and $\omega$
is found in appendix B.
Because the vevs of the supersymmetry transformation of matter fermions 
in the fundamental  representation
are proportional to those of
matter bosons in the fundamental representation,
they vanish.
On the other hand,
the vevs of the supersymmetry transformation of matter fermions 
in the adjoint representation
are proportional to those of
matter bosons in the adjoint representation
which are in non-Cartan directions,
and so they vanish.
Thus we have for all matter fermions
\begin{eqnarray}
\langle
\delta_{\eta}\Psi
\rangle=0~,~~~\Psi=\{
\kappa^u,
\psi^u,
\kappa^a,
\psi^a,
\kappa^u_i,
\psi^u_i,
\kappa^a_i,
\psi^a_i
\}~.
\end{eqnarray} 
Let us look at the mass terms of $\lambda^{\underline{\rm i}i}$ in (\ref{Lmass:QCD}).
Because of (\ref{vev:matter})
they reduce at the vacua to (\ref{lambda: mass}),
and thus $U^1{}_j\lambda^{\underline{\rm i}j}$ is massless.
This means in
the ordinary basis spanned by
 hermitian matrices $t_a$ of $U(N_c)$,
that
the overall $U(1)$ fermion $U^1{}_j\lambda^{0j}$ 
is the Nambu-Goldstone fermion
associated with the spontaneous partial supersymmetry breaking $\CN=2\to \CN=1$.

In summary,
we find that 
in the Coulomb phase $\langle\phi^{\underline{\rm i}}\rangle\neq 0$
the $\CN=2$ $U(N_c)$ QCD model $S_{\rm QCD}$
(\ref{S_QCD})
describes the spontaneous partial supersymmetry breaking 
$\CN=2\to\CN=1$.

\bigskip

\section*{Acknowledgments}
The authors thank
Keisuke Ohashi and 
Norisuke Sakai
for useful discussions.
This work is supported in part by the Grant-in-Aid for Scientific
Research (No.~16540262, No.~17540262 and No.~17540091) 
from the Ministry of Education,
Science and Culture, Japan.
Support from the 21 century COE program
``Constitution of wide-angle mathematical basis focused on knots"
is gratefully appreciated.
The preliminary version of this work was presented 
by M.S.
at the YITP workshop on
``String Theory and Quantum Field Theory"
in the Yukawa Institute
for Theoretical Physics, Kyoto University (19-23 August, 2005).
We wish to acknowledge the participants
for stimulating discussions.

\newpage
\appendix
\section{Notation}
The bar `` $\bar{~}$ '' means the
complex conjugation 
\begin{eqnarray}
\overline{f^i}&=&
\bar f_i~,\\
\overline{f_i}&=&
\overline{\epsilon_{ij}f^j}
=\epsilon_{ij}\bar f_j
=\epsilon_{ij}\epsilon_{jk}\bar f^k
=-\bar f^i~,\\
\overline{\xi^{ij}}&=&
\bar \xi_{ij}~,\\
\overline{\xi_{ij}}&=&
\overline{\epsilon_{ik}\epsilon_{jl}\xi^{kl}}
=
\epsilon_{ik}\epsilon_{jl}\bar\xi_{kl}
=
\epsilon_{ik}\epsilon_{jl}\epsilon_{kp}\epsilon_{lq}\bar\xi^{pq}
=
\bar\xi^{ij}
~.
\end{eqnarray}
The tilde `` $\widetilde{~} $ ''
is the analyticity preserving conjugation,
the product of the complex conjugation
`` $\bar{~}$ ''
 and the antipodal map `` $^\star$ '':
$(u^{+i})^\star=u^{-i}$ and $(u^{-i})^\star=-u^{+i}$
\begin{eqnarray}
\widetilde{u^\pm_i}&=&u^{\pm i}~,~~~
\widetilde{u^{\pm i}}=-u^{\pm}_i~,
\\
\widetilde{\theta^\pm}&=&\bar\theta^\pm~,~~~
\widetilde{\bar \theta^\pm}=-\theta^\pm~.
\end{eqnarray}

\section{$\CN=2$ supersymmetry transformation}

Under $\CN=2$ supersymmetry, coordinates transform as
\begin{eqnarray}
(x^m_A,\theta^{\pm}_\alpha, \bar\theta^{\pm\dot\alpha},u^\pm_i)
\to
(x^m_A-2i(\eta^i\sigma^m\bar\theta^++\theta^+\sigma^m\bar\eta^i)u^-_i,
\theta^{\pm}_\alpha+u^\pm_i\eta^i_\alpha, 
\bar\theta^{\pm\dot\alpha}+u^\pm_i\bar\eta^{i\dot\alpha},
u^\pm_i)~,~~
\end{eqnarray}
which is generated by the differential operator
\begin{eqnarray}
\eta Q+\bar\eta \bar Q~~~\mbox{with}~~~
\left\{
  \begin{array}{l}
Q_{\alpha i}=
-2i\sigma^m\bar\theta^+u^-_i\partial_m
+u^+_i\partial_{\theta^{+\alpha}}
+u^-_i\partial_{\theta^{-\alpha}}~       \\
\bar Q_{\dot\alpha i}=
-2i\theta^+\sigma^mu^-_i\partial_m
+u^+_i\partial_{\bar\theta^{+\dot\alpha}}
+u^-_i\partial_{\bar\theta^{-\dot\alpha}}       \\
  \end{array}
\right..
\end{eqnarray}
The supersymmetry transformation law of component fields
in $V^{++}$
is derived by
matching the components with
the appropriate power of $\theta^\pm$, $\bar\theta^\pm$
in 
$\delta_\eta V^{++}=(\eta Q+\bar\eta \bar Q)V^{++}$
where the left hand side means transformations on the component fields.
Because we are working in the WZ gauge, 
we gauge transform the resulting expression into the WZ gauge form.
Thus we examine
\begin{eqnarray}
\delta_\eta V^{++}=(\eta Q+\bar\eta \bar Q)V^{++}
-\delta_{g}V^{++}.~
\end{eqnarray}
Because the supersymmetry transformation
is an infinitesimal transformation,
it is enough to consider the infinitesimal gauge transformation
$\delta_{g}V^{++}=-\CD^{++}\lambda_g$.
Observe that
by choosing an analytic superfield $\lambda_g$ 
as
\begin{eqnarray}
\lambda_g&=&
F_g
+\theta^+\lambda^-_g
+\bar\theta^+\bar\kappa^-_g
+(\theta^+)^2M^{--}_g
+(\bar\theta^+)^2N^{--}_g
+i\theta^+\sigma^m\bar\theta^+ A_{gm}^{--}
\nonumber\\&&
+(\theta^+)^2\bar\theta^+\bar\chi^{(-3)}_g
+(\bar\theta^+)^2\theta^+\lambda^{(-3)}_g
+(\theta^+)^4P^{(-4)}_g
\label{lambda_g}
\end{eqnarray}
with
\begin{eqnarray}
F_g&=&P^{(-3)}_g=0~,\\
\lambda^-_{g\alpha}&=&
2i(\sqrt{2}\eta^i\bar\phi+\sigma^m\bar\eta^iv_m)u_i^-~,\\
\bar\kappa_g^{-\dot\alpha}&=&
-2i(\sqrt{2}\bar\eta^i\phi+\bar\sigma^m\eta^iv_m)u_i^-~,\\
M^{--}_g&=&
2\bar\eta^i\bar\lambda^ju^-_iu^-_j~,\\
N^{--}_g&=&
-2\eta^i\lambda^ju^-_iu^-_j~,\\
A_{gm}^{--}&=&
2i(\eta^i\sigma_m\bar\lambda^j-\lambda^i\sigma_m\bar\eta^j)u^-_{i}u^-_{j}~,\\
\lambda^{(-3)}_g&=&
-2\eta^iD^{jk}u^-_{(i}u^-_ju^-_{k)}~,\\
\bar\chi^{(-3)}_g&=&
-2\bar\eta^iD^{jk}u^-_{(i}u^-_ju^-_{k)}~,
\end{eqnarray}
the $\CN=2$ supersymmetry transformation law
is obtained as
\begin{eqnarray}
\delta_\eta\phi&=&
-i\sqrt{2}\epsilon_{ij}\eta^i\lambda^j~,\\
\delta_\eta\bar\phi&=&
-i\sqrt{2}\epsilon_{ij}\bar\eta^i\bar\lambda^j~,\\
\delta_\eta v_m&=&
i\epsilon_{ij}(
\eta^i\sigma_m\bar\lambda^j+\lambda^i\sigma_m\bar\eta^j
)
~,\\
\delta_\eta\lambda^i_\alpha&=&
\frac{1}{2}\sigma^m\bar\sigma^n\eta^iv_{mn}
+\sqrt{2}\sigma^m\bar\eta^i\CD_m\phi
-i\eta^i[\phi,\bar\phi]
+
D^{i}{}_j\eta^j~,\\
\delta_\eta\bar\lambda^{i\dot\alpha}&=&
\frac{1}{2}\bar\sigma^m\sigma^n\bar\eta^iv_{mn}
-\sqrt{2}\bar\sigma^m\eta^i\CD_m\bar\phi
+i\bar\eta^i[\bar\phi,\phi]
-D^{i}{}_j
\bar\eta^j~,\\
\delta_\eta D^{ij}&=&
-2i\eta^i\sigma^m\CD_m\bar\lambda^j
+2i\CD_m\lambda^i\sigma^m\bar\eta^j
+2\sqrt{2}\bar\eta^i[\bar\lambda^j,\phi]
+2\sqrt{2}\eta^i[\lambda^j,\bar\phi]~.
\end{eqnarray}

Because the $U(N_c)$ gauged action for $q^+$ hypermultiplets
is invariant under the infinitesimal gauge transformation,
$\delta_g q^+=i\lambda q^+$ and $\delta_gV^{++}=-\CD^{++}\lambda$
with an analytic superfield $\lambda$,
the supersymmetry transformation of $q^+$ must be followed
by the gauge transformation with $\lambda\equiv\lambda_g$.
Thus the supersymmetry transformation of the component fields in
the $q^+$ hypermultiplet can be read off from
\begin{eqnarray}
\delta_\eta q^{+}=(\eta Q+\bar\eta \bar Q)q^{+}
-\delta_{g}q^{+}~
\end{eqnarray}
with $\delta_{g}q^{+}=i\lambda_g q^+$
as
\begin{eqnarray}
\delta_\eta f^i&=&
\eta^i\psi+\bar\eta^i\bar\kappa~,\\
\delta_\eta \psi_\alpha&=&
2i(\sigma^m\bar\eta^i)_\alpha\CD_mf_i
-2\sqrt{2}\eta^i_\alpha\bar\phi f_i~,\\
\delta_\eta \bar\kappa_{\dot\alpha}&=&
2i(\eta^i\sigma^m)_{\dot\alpha}\CD_mf_i
+2\sqrt{2}\bar\eta^i_{\dot\alpha}\phi f_i~.
\end{eqnarray}
Similarly by examining
\begin{eqnarray}
\delta_\eta \boldsymbol{\omega}=(\eta Q+\bar\eta \bar Q)\boldsymbol{\omega}
-\delta_{g}\boldsymbol{\omega},~
\end{eqnarray}
with $\delta_{g}\boldsymbol{\omega}=i\lambda_g \boldsymbol{\omega}$,
the supersymmetry transformation of the component fields
in $\boldsymbol{\omega}$ hypermultiplet is obtained as
\begin{eqnarray}
\delta_\eta\omega&=&
\sqrt{2}(
\eta^i\psi_i
+\bar\eta^i\bar\kappa_i
)
~,\\
\delta_\eta\omega^{ij}&=&
2(
\eta^{(i}\psi^{j)}
+\bar\eta^{(i}\bar\psi^{j)}
)
~,\\
\delta_\eta\psi^i_\alpha&=&
-\sqrt{2}i(\sigma^m\bar\eta^i)_\alpha\CD_m\omega
+2\eta^i_\alpha\bar\phi\omega
+2i(\sigma^m\bar\eta^j)_\alpha\CD_m\omega_j{}^i
-2\sqrt{2}\eta^j_\alpha\bar\phi\omega_j{}^i
~,\\
\delta_\eta\bar\kappa^{i}_{\dot\alpha}&=&
-\sqrt{2}i(\eta^i\sigma^m)_{\dot\alpha}\CD_m\omega
-2\bar\eta^{i}_{\dot\alpha}\phi\omega
+2i(\eta^j\sigma^m)_{\dot\alpha}\CD_m\omega_j{}^i
+2\sqrt{2}\bar\eta^{j}_{\dot\alpha}\phi\omega_j{}^i
~.
\end{eqnarray}

In presence of the magnetic FI term,
the auxiliary field $D^{ij}$
is shifted by an imaginary constant.
The $U(N_c)$ gauge model (\ref{S_YM})
and the $U(N_c)$ QCD model (\ref{S_QCD})
are invariant under the above
 supersymmetry transformations
with the replacement $D\to \boldsymbol{D}$.

\section{symplectic covariant form of $S^{\rm gauged}_{q}$}
Here, we give the symplectic covariant form of $S^{\rm gauged}_{q}$.
The $q^{+}$ can be written as
\begin{eqnarray}
&&q^{+M }=
F^{+ M}(x_A,u)
+\theta^{+}\Psi^M (x_A,u)
+\bar\theta^+\bar\Psi^M (x_A,u)
+(\theta^+)^2M^{- M}(x_A,u)
\nonumber\\&&
+(\bar\theta^+)^2N^{- M}(x_A,u)
+i\theta^+\sigma^m\bar\theta^+A_m^{- M}(x_A,u)
+(\theta^+)^2\bar\theta^+\bar\chi^{(-2) M}(x_A,u)
\nonumber\\&&
+(\bar\theta^+)^2\theta^+\chi^{(-2) M}(x_A,u)
+(\theta^+)^2(\bar\theta^+)^2P^{(-3) M}(x_A,u)~
\nonumber
\end{eqnarray}
where
$M$ runs from $1$ to $2N_c$ for $U(N_c)$ fundamental $q^+$
while from $0$ to $2N_c^2-1$ for $U(N_c)$ adjoint $q^+$.
$q^+_M$ is related to $q^{+M}$ by
$
\widetilde{q^+_M}\equiv q^{+M}=\Omega^{MN}q^{+}_{N}~,
$
where $\Omega^{MN}=\Omega^{NM}$ is the invariant tensor of the symplectic group,
$Sp(N_c)$ for fundamental $q^+$
or $Sp(N_c^2)$ for adjoint $q^+$.
The gauged action is
\begin{eqnarray}
S^{\rm gauged}_q&=&
\frac{1}{2}\intd ud\zeta^{(-4)}\left[
q^+_M \CD^{++}q^{+M}
\right]~,
\\
\CD^{++}q^{+M}&=&D^{++}q^{+M}+i\boldsymbol{V}^{++M}{}_N q^{+N}~,~~~
\boldsymbol{V}^{++M}{}_N=V^{++a}\boldsymbol{T}_a~,~~~
\boldsymbol{T}_a
=
\left(
  \begin{array}{cc}
   -T_a^{T}    &0    \\
  0     &T_a    \\
  \end{array}
\right)~.\nonumber
\end{eqnarray}
The fundamental or adjoint representation of the gauge group
is embedded in the symplectic matrix, 
$Sp(N_c)$
or $Sp(N_c^2)$,
respectively.
This action is invariant under $U(N_c)$ gauge transformation
$
\delta q^{+M}=\epsilon^a\lambda_a^{+M}$
with
$
\lambda_a^{+M}=i(\boldsymbol{T}_a)^M{}_N q^{+N}~.
$

\newpage

\end{document}